\documentclass[aps,onecolumn,prl]{revtex4}
\usepackage{latexsym, verbatim}
\usepackage{amsmath}
\usepackage{mathrsfs}
\usepackage{amsthm}
\usepackage{amssymb}
\usepackage{dsfont}
\usepackage{fancybox}
\usepackage{bm}
\usepackage{fancyhdr}
\usepackage{subfigure}
\usepackage{graphicx}
\usepackage{color}

\begin{document}

\title{\bf Asymmetric Wave Propagation Through 
Nonlinear PT-symmetric
Oligomers}

\author{J. D'Ambroise}
\affiliation{Department of Mathematics,
Bard College,
Annandale, NY 12504, USA} 

\author{P.G.~Kevrekidis}
\affiliation{Department of Mathematics and Statistics, University of Massachusetts,
Amherst, Massachusetts 01003-4515, USA}

\author{S. Lepri}
\affiliation{CNR-Consiglio Nazionale delle Ricerche, Istituto dei Sistemi
Complessi, via Madonna del piano 10, I-50019 Sesto Fiorentino, Italy} 

\begin{abstract}
In the present paper, we consider nonlinear PT-symmetric dimers and trimers 
(more generally, oligomers) embedded within a linear Schr{\"o}dinger
lattice. We examine the stationary states of such chains in the form
of plane waves, and analytically compute their reflection and transmission
coefficients through the nonlinear PT symmetric oligomer, as well as the
corresponding rectification factors which clearly illustrate the asymmetry
between left and right propagation in such systems. We examine
not only the existence but also the dynamical stability of the plane wave 
states and interestingly find them to be unstable
{except in the vicinity of the linear limit}. Lastly, we
generalize our numerical considerations to the more physically relevant
case of Gaussian initial wavepackets and confirm that the asymmetry in
the transmission properties persists in the case of such wavepackets,
as well. { The role of potential asymmetries in the nonlinearity or
in the gain/loss pattern is also considered.}
\end{abstract} 

\maketitle

\section{Introduction} 

Over the last fifteen years, and ever since its original proposal
by Bender and co-workers~\cite{bend}, the study of PT-symmetric
Hamiltonian systems has become a focal point for numerous investigations
at the interface between theoretical physics and applied mathematics.
The fundamental appeal of such systems is that they respect key 
physical symmetries, namely the Parity (P)
and Time-reversal (T) (but not necessarily the stronger constraint
of the Hamiltonian being Hermitian), 
thus providing an intriguing alternative set of 
Hamiltonians with potentially real eigenvalues. In the context that
we will examine below and for standard   Schr{\"o}dinger 
Hamiltonians with a complex
potential $V$, the above constraints of 
PT symmetry amount to the potential satisfying
the condition  $V(x)=V^{\star}(-x)$. 

While this field commenced as, arguably, a mathematical curiosity
associated with the foundations of quantum mechanics, a number of
major
developments from the point of view of applications arose recently.
{ Initially, it was realized that electromagnetic settings could
provide suitable experimental systems for the realization
of linear PT-symmetric Hamiltonians \cite{muga}.}
However, a considerable volume of developments materialized due to the work of
Christodoulides and co-workers~\cite{christo1} who realized that nonlinear
optics (and the synthetic systems that can be engineered therein)
may present a fertile playground for the experimental implementation
of such PT-symmetric dynamics. Furthermore, the added feature of
nonlinearity typically present in such settings initiated the
consideration of the effects of such PT-symmetric potentials on
the nonlinear waves (such as bright or gap solitons) that may
arise therein. Subsequently, the first realizations of PT-symmetry
arose both in the
nonlinear optics of waveguide couplers (i.e., either 
two waveguides with and without loss~\cite{salamo} -- the so-called 
passive PT-- or in the more ``standard'' case of one waveguide
with gain and one with loss~\cite{kip}) and also in that of electronic analogs
thereof~\cite{tsampikos_recent}. This progress and the perspectives
that it enables towards future developments have, in turn, fueled
a considerable volume of further theoretical studies. These are 
concerned both
with the realm of models with PT-symmetric 
potentials in the presence of 
nonlinearity~\cite{kot1,sukh1,kot2,grae1,grae2,kot3,pgk,dmitriev1,dmitriev2} 
and even with that of models where gain-loss contributions of a balanced
form  appear in front of the nonlinear
term~\cite{miron,konorecent,konorecent2}.

Another theme that has received considerable attention recently
concerns the study of asymmetric (i.e., non-reciprocal) wave propagation
in the context of various applications. Among the 
first examples discussed in the literature is the 
asymmetric phonon transmission through a nonlinear interface 
layer between two very dissimilar crystals \cite{kosevich}.
Later on, the theoretical
proposition of a thermal diode~\cite{l1} induced relevant experimental
realizations in~\cite{l3}; similarly, an optical diode~\cite{l5}
was theoretically proposed~\cite{l7} and experimentally realized~\cite{l9}.
More recently similar proposals have been presented in left-handed
metamaterials~\cite{l10} and have, in fact, been experimentally
realized by different groups also in the context of acoustic waves
in granular systems~\cite{chiara1,others}. In fact, recently this
theme has been examined in the context of PT periodic structures
which have been shown~\cite{kot3} to act as unidirectional invisible
media (at least for sufficiently small extents of the
periodic structure~\cite{longhi}) with transmission coefficients
and phases identical to the ones in the absence of the PT-structure.

In the present work, we adopt a different perspective to that of
these recent works and in fact one closer to the considerations of
~\cite{lepri}. The latter work considered a linear lattice in the
presence of a set of $N$ (embedded within the linear lattice) nonlinear
elements. 
It was then shown that an asymmetry within the linear or nonlinear
(still Hamiltonian though) properties of these elements would lead
to an asymmetry of the relevant transmission of a plane
wave through the chain. This was then extended to the more physically
realistic case of a Gaussian wavepacket whose asymmetry was examined
between incidence (on the asymmetric nonlinear region) from the left
and from the right.

Here, we consider these notions of asymmetric wave
propagation but for PT-symmetric oligomers (namely dimers, trimers,
quadrimers, etc.; see also the earlier study of such oligomers~\cite{pgk}
which has motivated further recent studies such as the examination of
quadrimers in~\cite{vvk}). In particular, 
we examine both the
similarities and the differences with the above picture. In particular,
we start our considerations by examining the transmission of plane waves 
through the chain. For such states, we analytically identify their
reflection and transmission coefficients and obtain the corresponding
transmittivity. We find some fundamental differences here in that
such a quantity may exceed unity due to the presence of gain.
Furthermore, we identify another fundamental difference in that
beyond a particular gain strength a supercritical amount of ``mass''
may be collected at the gain site that may in turn lead to indefinite
growth
on this site. Nevertheless, for PT-coefficient strengths which yield
meaningful transmittivities, a rectification factor is computed
and clearly evidences the asymmetry between left and right propagation.
Another aspect of the relevant plane wave solutions that we clarify
concerns their generic dynamical instability, which we quantify
and illustrate the (again, focusing) dynamical manifestation thereof.
Lastly, we generalize our considerations (numerically) to the prototypical
physically relevant case of an incident Gaussian wavepacket.
We showcase that in that case as well, the transmission result
differs between left and right such wavepackets impinging on the 
nonlinear PT-symmetric lattice region.

{ It should be noted that such linear lattices with nonlinear
``impurities'' have been considered in the past in electronic
settings (where the nonlinearity characterizes the strong
interaction with local vibrations at the impurity site)~\cite{tsir1}.
They have also been examined in magnetic settings to describe tunneling
through a magnetic impurity connected to two perfect leads in the
presence of a magnetic field~\cite{moli} and have even been generalized
to higher dimensions~\cite{moli2}. Here, we envision an optical
setting where two types of waveguides are used such that one of
them is linear for the considered propagation. A recent example of such binary
waveguides (yet in an alternating fashion so as to form a periodic
linear-nonlinear waveguide array) appears e.g. in~\cite{efrem},
where narrower waveguides are highly nonlinear whereas wider ones are
almost linear. It is, however, straightforward to mold the relevant
geometry of these ribs (appearing e.g. in Fig. 1 of~\cite{efrem})
in order to construct the lattice of interest herein. Then, the gain
and loss can additionally be imparted to the lattice according to
the prescription of~\cite{kip}.}

Our presentation is structured as follows. In section II, we will
give an overview of the relevant theory (some of the pertinent, 
details are relegated to an appendix). In section III,
the analytical results are illustrated through numerical computations
of both the transmittivities and rectification factors (both for
the plane wave structures and for the Gaussian wavepackets) and
of the existence, spectral stability and nonlinear dynamics of the 
plane waves. Finally, in section IV, we summarize our findings
and present some directions for future studies.

\section{Theoretical Analysis}

\subsection{Stationary DNLS}
\label{secstatDNLS}

Starting our considerations with the existence
of stationary solutions of the infinite chain of interest, we examine
the set of algebraic equations
\begin{equation} \omega \psi_n = V_n \psi_n - \psi_{n+1}-\psi_{n-1}+\alpha_n |\psi_n|^2\psi_n\label{eq: statDNLS}\end{equation}
on a one-dimensional lattice for $\psi_n\in\mathds{C}$ and $\omega\in\mathds{R}$.  The parameters $V_n\in\mathds{C}$ and $\alpha_n\in\mathds{R}$ are zero everywhere except for $1\leq n \leq N$ so that the wave propagates freely outside of a finite region containing the nonlinearity and the (intended to be
PT-symmetric, hence generally complex) linear potential.  
Rearranging (\ref{eq: statDNLS}) one obtains the so-called backward 
transfer map
\begin{equation}\psi_{n-1}=-\psi_{n+1}+(V_n-\omega+\alpha_n |\psi_n|^2)\psi_n \label{eq: BXFerMap}\end{equation}
from which solutions to (\ref{eq: statDNLS}) can be constructed from knowledge of $\psi_n$ at two nodes.

The class of solutions whose (transmission and reflection) properties
will be thereotically analyzed consists of  plane waves of the form
\begin{equation}  \psi_n=\left\{ 
\begin{array}{ll}
R_0e^{ik_0n}+Re^{-ik_0n} & n\leq 1\\
Te^{ik_0n} & n\geq N
\end{array}
\right.
\label{eq: FormOfPsi_n}  \end{equation}
with $R_0, R, T\in\mathds{C}$ representing the incident, reflected and transmitted amplitudes, respectively { and $k_0 \ge 0$ is the wavenumber}.   
Note that outside of the nonlinear region, for $n<1$ and $n>N$, (\ref{eq:  FormOfPsi_n}) satisfies (\ref{eq: statDNLS}) for any $R_0, R$ and $T$ if $\omega=-2 \cos(k_0)$.  
Also directly from (\ref{eq: FormOfPsi_n}) [as applied
to sites with $n=0$ and $n=1$], we have 
\begin{equation}R_0=\frac{e^{-ik_0}\psi_0-\psi_1}{e^{-ik_0}-e^{ik_0}} \qquad \mbox{ and } \qquad  R=\frac{e^{ik_0}\psi_0-\psi_1}{e^{ik_0}-e^{-ik_0}}.  \label{eq: R0RwrtPsi}\end{equation}
Thus, for any fixed values of $k_0$ and $T$, $\psi_0$ and $\psi_1$ can be calculated by applying (\ref{eq: BXFerMap}) iteratively starting with $\psi_N=Te^{ik_0N}$ and $\psi_{N+1}=Te^{ik_0(N+1)}$ from (\ref{eq:  FormOfPsi_n}).  Then,  
(\ref{eq:  R0RwrtPsi})  gives  the appropriate values of  $R_0$ and $R$ so that (\ref{eq: statDNLS}) is satisfied at all nodes. { Such a
procedure of finding the input as a function
of the output (sometimes referred to as a ``fixed output problem'' \cite{knapp}) is 
necessary to deal with the multistability which is often found for nonlinear 
problems, including our case (at least for a subset of parameters)}.

% if l=1 then the backwards transfer map writes phi_{N-1} in terms of phi_{N+1} and phi_{N}, delta_N
% if l=2 then phi_{N-2} is in terms of phi_{N} and phi_{N-1}, delta_{N-1}
% if l=3 then phi_{N-3} is in terms of phi_{N-1} and phi_{N-2}, delta_{N-2}
% if l=4 then phi_{N-4} is in terms of phi_{N-2} and phi_{N-3}, delta_{N-3}
% if l=5 then phi_{N-5} is in terms of phi_{N-3} and phi_{N-4}, delta_{N-4}
% \vdots
% if l=N-1 then phi_{1} is in terms of phi_{3} and phi_{2}, delta_{2}
% if l=N then phi_{0} is in terms of phi_{2} and phi_{1}, delta_{1}

%okay, so it is delta_1 and delta_2 that are NOT equal to -\omega
%For all other nodes we have delta_j=-omega

%if l=0 and N=2 then we get 
% phi_2 = - phi_{4} +delta_{3}phi_{3}

For convenience we write the backward transfer map with $n=N-l+1$ and in terms of $\Psi_n$ for $\psi_n\stackrel{def.}{=}Te^{ik_0N}\Psi_n$.  This gives
for our infinite lattice:
\begin{equation}
\Psi_{N-l}=-\Psi_{N-l+2}+\delta_{N-l+1}\Psi_{N-l+1}\label{eq: alg}
\end{equation}
for $\delta_j\stackrel{def.}{=}V_j-\omega+\alpha_j|T|^2|\Psi_j|^2$.  For example, applying (\ref{eq: alg}) with $N=2$ and  $\Psi_3 = e^{ik_0}, \Psi_2 = 1$  gives 
\begin{eqnarray}\label{eq: N=2alg}
\delta_2 &=& V_2-\omega+\alpha_2|T|^2 \\
\Psi_1&=& -e^{ik_0}+\delta_2\nonumber\\
\delta_1 &=& V_1-\omega+\alpha_1|T|^2|\delta_2-e^{ik_0}|^2\nonumber\\
\Psi_0 &=& -1 + \delta_1(\delta_2-e^{ik_0}).\nonumber
\end{eqnarray}
Finally 
by (\ref{eq: R0RwrtPsi}) and (\ref{eq: N=2alg}) we have 
\begin{equation} 
R_0 =\frac{Te^{ik_0}}{ e^{-ik_0}-e^{ik_0}} \left(   -1+(\delta_1-e^{ik_0})(\delta_2-e^{ik_0}) \right)
\end{equation}
and the corresponding transmission coefficient $t\stackrel{def.}{=} |T|^2/|R_0|^2$ is then
\begin{equation}
t=\left| \frac{e^{ik_0}-e^{-ik_0}}{ 1+(\delta_1-e^{ik_0})(e^{ik_0}-\delta_2)  }
\right|^2  .
\end{equation}
In the linear case ($\alpha_1=\alpha_2=0$), it is immediately seen
that $t$ is the same for waves
coming from the left or right side, independently of $V_n$,
as prescribed by the reciprocity theorem. 
{ For a more detailed discussion of the linear case 
see e.g. Ref.~\cite{lindquist}. }

It should be noted here that although the relevant calculation was
presented for $N=2$, it can be performed for arbitrary values of $N$
(naturally, the complexity of the intermediate steps is increased,
the higher the value of $N$). 
%For reasons of clarity of exposition
%but also of completeness, the relevant presentation has been given
Some of the relevant details for larger $N$ have been provided 
in the Appendix. We should also note that in comparison to the earlier
work of~\cite{lepri}, there are fundamental similarities in the approach
but also important differences in the nature of the results since
our quantities $\delta_i$ (corresponding to the $\nu$ and $\delta$,
respectively in~\cite{lepri}) are now, in principle, complex rather
than purely real.
Let us also indicate here that to quantify asymmetric
propagation we will use the definition of
a rectification factor $f$ in the form:
\begin{eqnarray}
f = \frac{t(k_0,T) - t(-k_0, T)}{t(k_0,T) + t(-k_0, T)},
\label{rectif}
\end{eqnarray}
{ where we adopt the convention that $-k_0$ denotes 
right-incoming solutions with wavenumber $k_0$.
Non-vanishing values} of $f$ in the range $[-1,1]$ are measures of the
asymmetry of transmission in the system (symmetric transmission is
tantamount to $f=0$).

\subsection{Time Propagation}
\label{secTDDNLS}

We also briefly touch upon the tools that we will use towards
the consideration of the dynamical evolution phenomena within
our nonlinear Schr{\"o}dinger type chains
\begin{equation} \label{eq: TDDNLS}i\dot\phi_n(t)-V_n\phi_n(t)+\phi_{n+1}(t)+\phi_{n-1}(t)=\alpha_n|\phi_n(t)|^2\phi_n(t). \end{equation}
In particular, in addition to direct numerical integration of
Eq.~(\ref{eq: TDDNLS}), so as to monitor the dynamical evolution 
of the solutions, we will use a spectral stability analysis of
stationary states (of the form $\psi_n e^{-i \omega t}$ discussed
in the previous section) according to
\begin{equation} \phi_n(t)=\rho_n(t)+\epsilon p_n(t).  \end{equation}
Here, $\rho_n(t)=\psi_n e^{-i \omega t}$ and $p_n(t)=e^{-i \omega t}  \left( a_ne^{i\nu t}+b_n e^{-i\nu^* t} \right)$ for $\omega \in\mathds{R}$ and $a_n, b_n,\nu \in\mathds{C}$.  $\rho_n$  is assumed to be a standing wave solution of (\ref{eq: TDDNLS}) so that the resulting order-$\epsilon$ equation  for $p_n(t)$ is 
\begin{equation}\label{eq: pneqn}
 i\dot{p}_n-V_np_n+p_{n+1}+p_{n-1} + \omega p_n= \alpha_n\left( 2p_n|\rho_n|^2+\rho_n^2{p_n^*} \right).
 \end{equation}
%Using the definition of $p_n$ and equating the coefficients of $e^{i(\Lambda+\nu)t}, e^{i(\Lambda-\nu^*)t}$ results in the matrix system
The ensuing linear stability equations will  yield the eigen-problem
of the form:
 \begin{equation}\label{eq: matrix}
 \nu \left(  \begin{array}{c} a_n\\ b_n^* \end{array}  \right)
 =
 \left( \begin{array}{cc} F_1 & F_2\\ F_3 & F_4 \end{array} \right)
\left(  \begin{array}{c} a_n\\ b_n^* \end{array}  \right)
\end{equation}
for
\begin{eqnarray}\label{eq: Fs}
F_1 &=& diag(\omega-V_n-2\alpha_n|\psi_n|^2)+G\\
F_2 &=& diag(-\alpha_n\psi_n^2)\nonumber\\
F_3 &=& diag(\alpha_n(\psi_n^*)^2)\nonumber\\
F_4 &=& diag(-\omega+V_n^*+2\alpha_n|\psi_n|^2)-G,\nonumber
\end{eqnarray}
where $G$ is a sparse matrix with ones on the superdiagonal and 
the subdiagonal.  Note that in (\ref{eq: matrix}) 
it is now convenient to think of $a_n$ and $b_n$ as column vectors.  Given a stationary solution $\rho_n$ and values of $V_n, \alpha_n$ which encode the nonlinearity for $1\leq n\leq N$, one then calculates the eigenvalues $\nu$ in (\ref{eq: matrix}).  If $\nu$ has a { negative} 
imaginary part this indicates that the perturbed solution $\phi_n(t)$ is unstable, as is easily seen by the form of $p_n(t)$ specified above.
In practice, one diagonalizes a finite truncation of the matrix 
in (\ref{eq: matrix}), ensuring that the relevant eigenvalues
are not affected by the truncation error. 
 
\section{Numerical Computations}

\begin{figure}[tbp]
\begin{center}
\includegraphics[width=8cm,angle=0,clip]{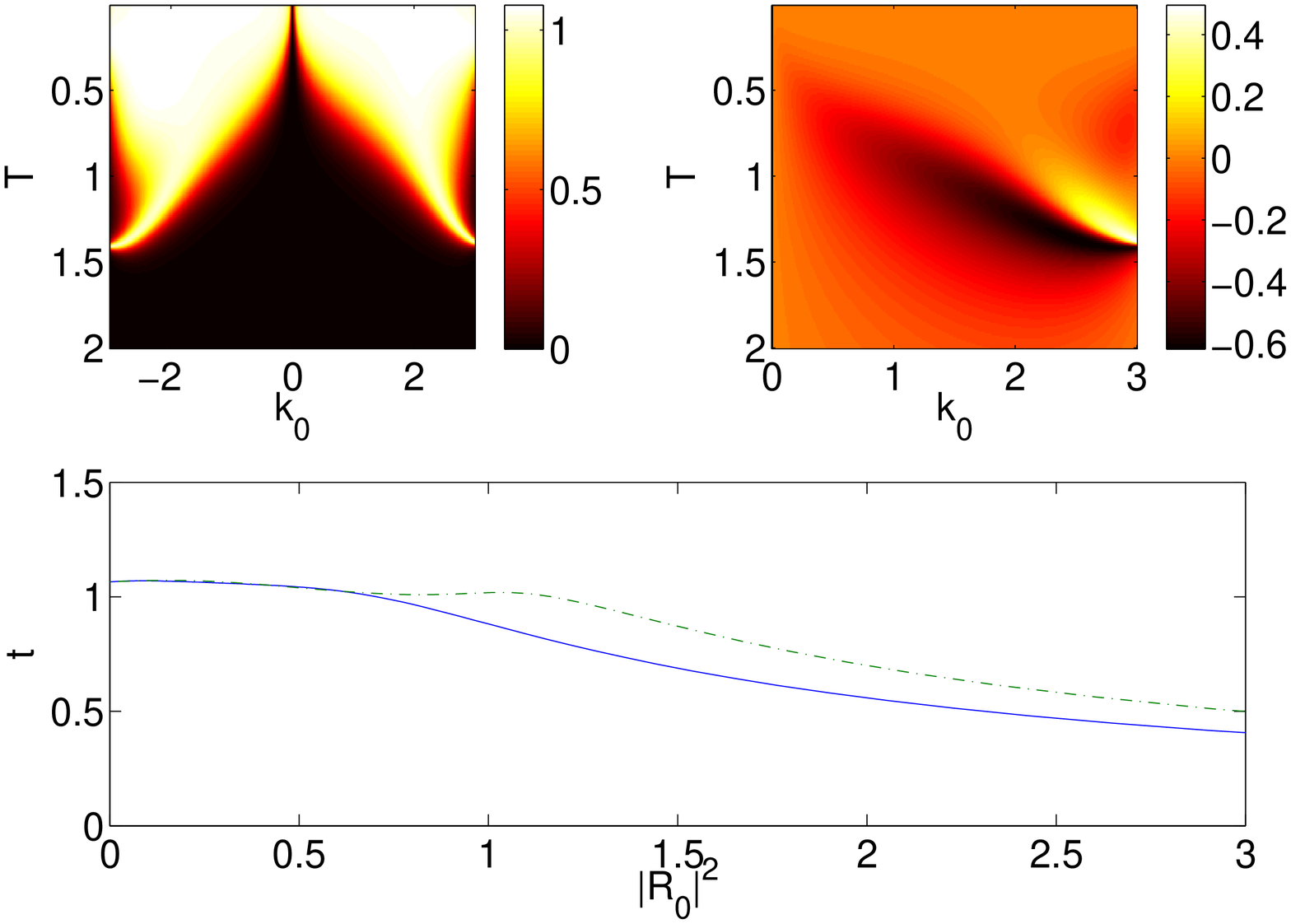}
\includegraphics[width=8cm,angle=0,clip]{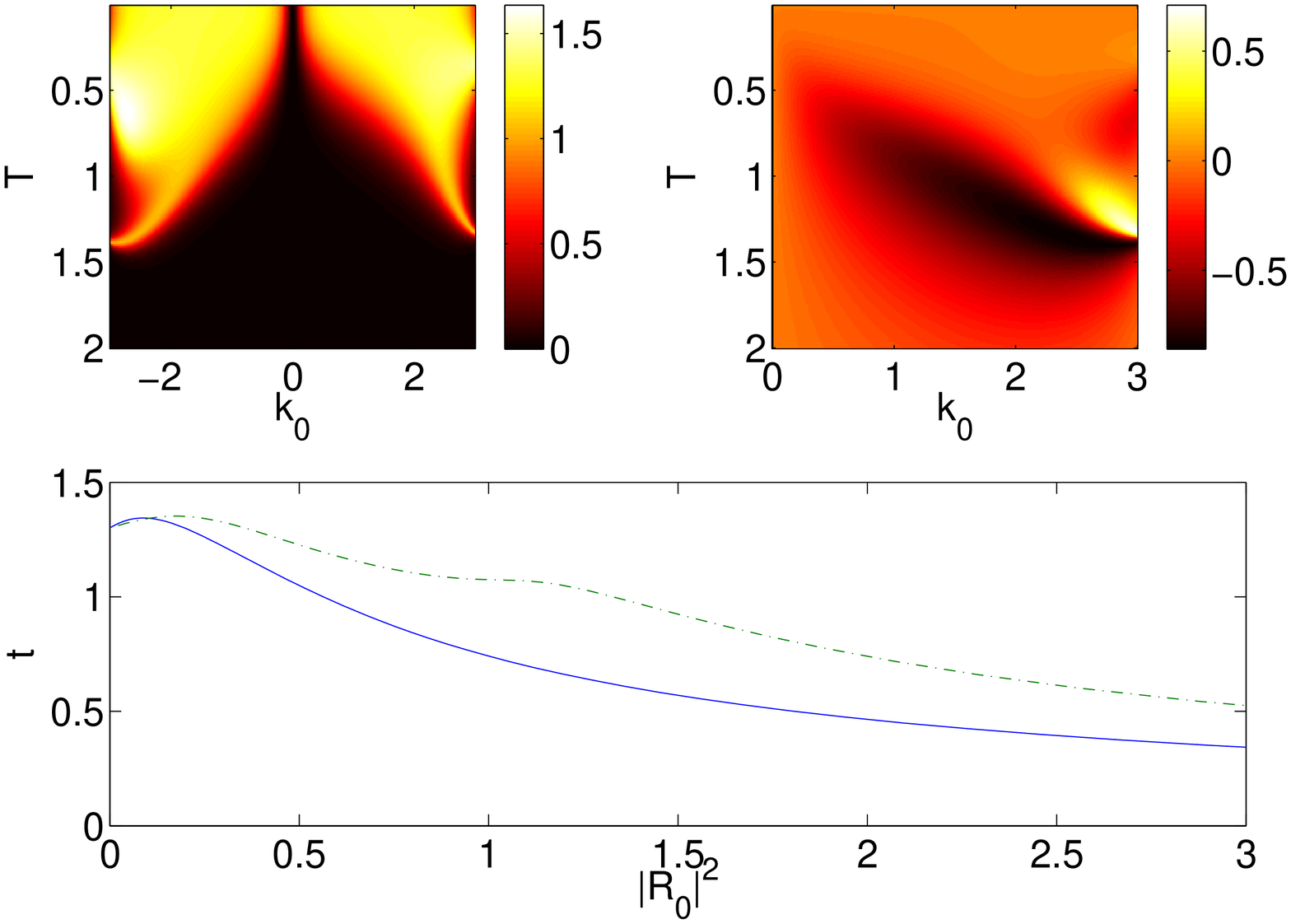}\\
%\epsfig{figure=PT_N2_g025.eps,height=2in,angle=0}
%\epsfig{figure=PT_N2_g05.eps,height=2in,angle=0}
%\epsfig{figure=../scripts1/PT_N2_g11.eps,height=2in,angle=0}
%\label{fig0}
\end{center}
\caption{The figure presents the case of a PT-symmetric dimer
with $N=2$, $\alpha_{1,2}=1$, while $V_1=i \gamma$ and $V_2=-i \gamma$.
The left set of panels corresponds to the case of $\gamma=0.25$,
while the right set of panels to the case of $\gamma=0.5$.
Each set contains a contour plot of $t(k_0,T)$ (top left),
a contour plot of $f(k_0,T)$ (top right) and a typical example of the
dependence of $t$ for $k_0=2$ (solid lines) and $k_0=-2$ (dashed lines), so as to illustrate
the asymmetry between the propagation for left and right incident
{ waves (bottom panels). 
In the latter the dependence of $t$ is given as a function of
$|R_0|^2$.}}

\label{fig0}
\end{figure}

\begin{figure}[tbp]
\begin{center}
\includegraphics[width=8cm,angle=0,clip]{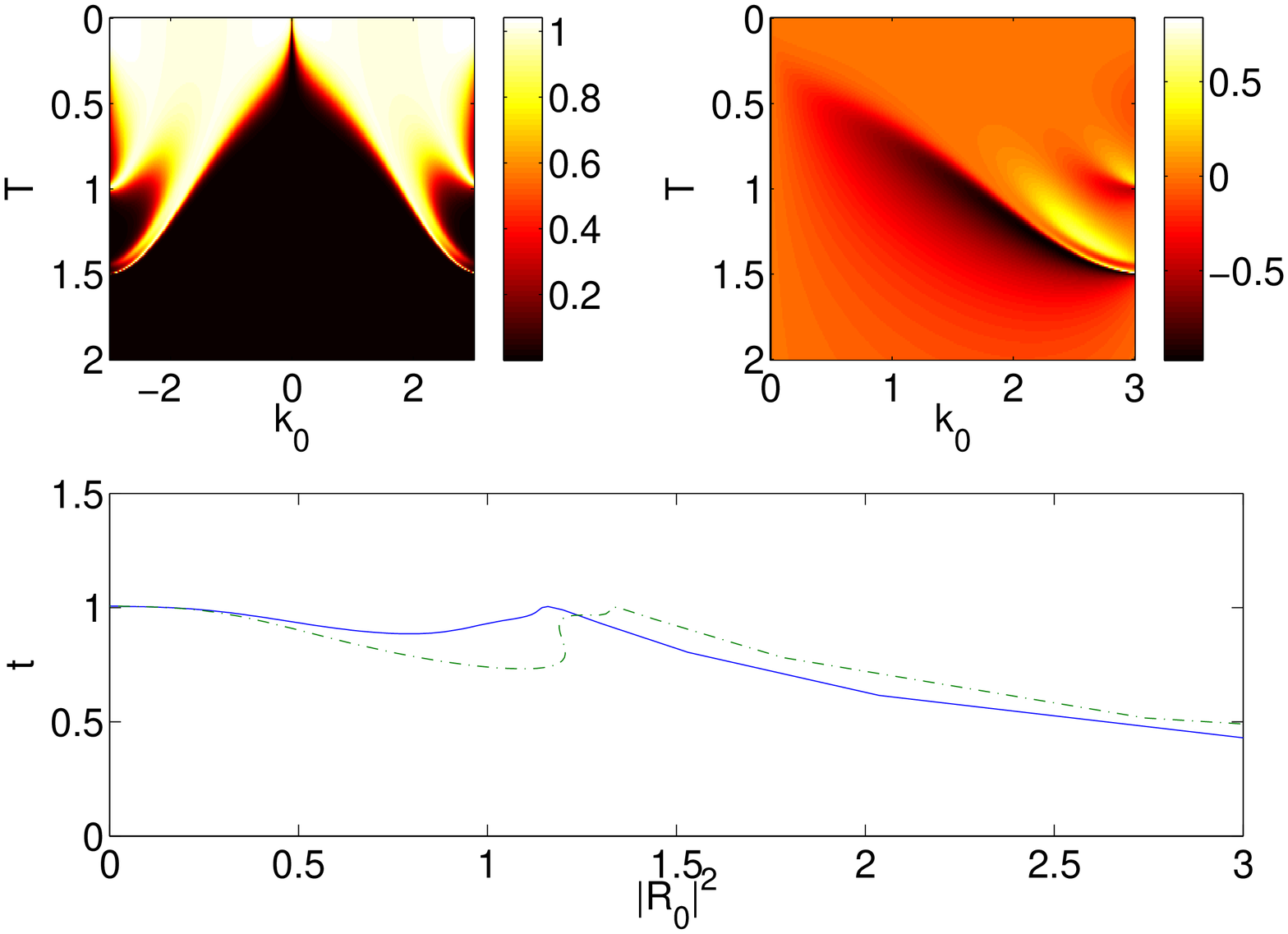}
\includegraphics[width=8cm,angle=0,clip]{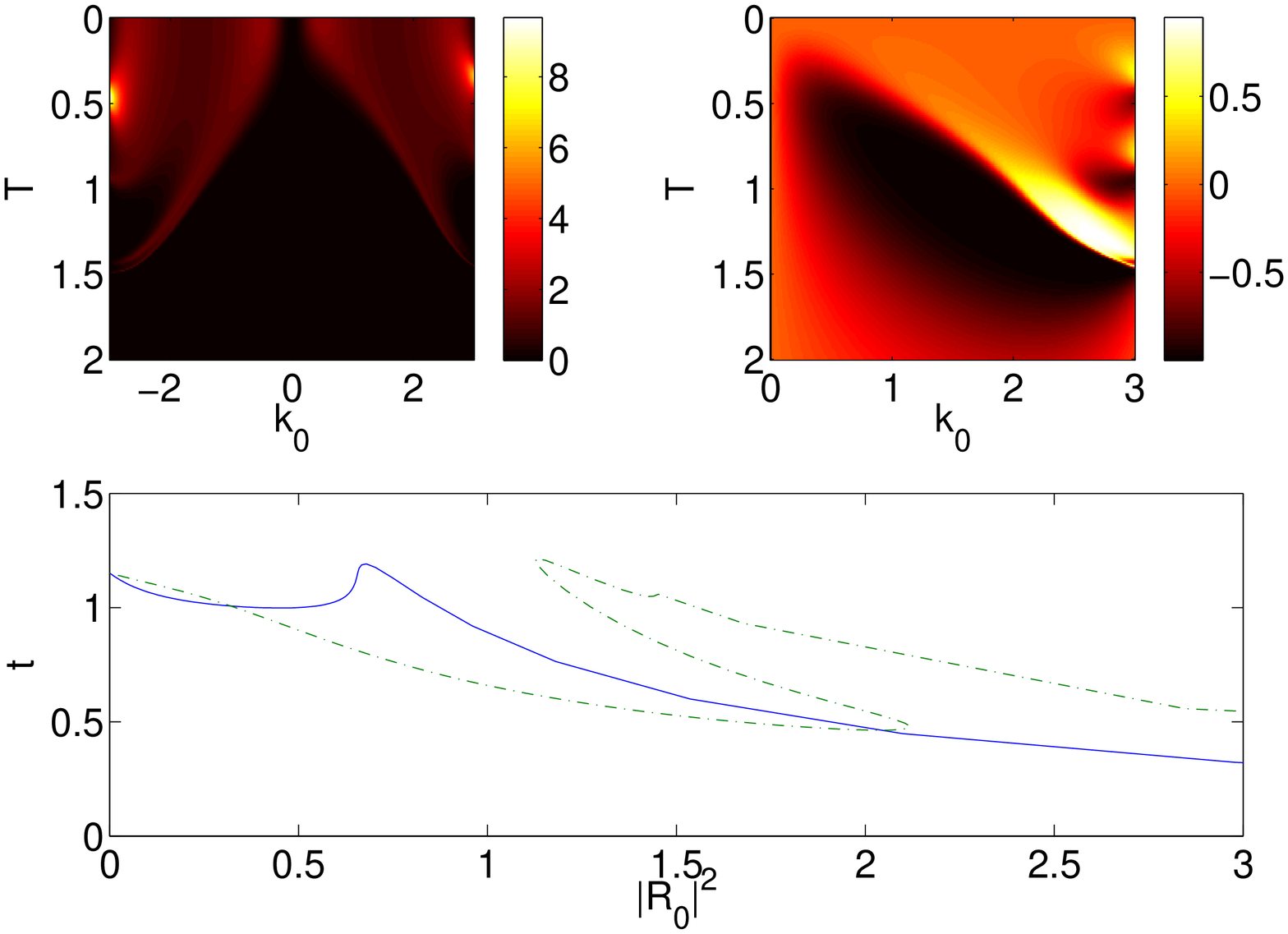}
\end{center}
%\label{fig1}
\caption{Same as the figure above but now for the trimer case
of $N=3$, $\alpha_{1,2,3}=1$, while $V_1=i \gamma$, $V_2=0$
and $V_3=-i \gamma$. The specific values of $\gamma$ for the panels
given are $\gamma=0.1$ and $\gamma=0.45$. }
\label{fig1}
\end{figure}

\begin{figure}[tbp]
\begin{center}
\includegraphics[width=8cm,angle=0,clip]{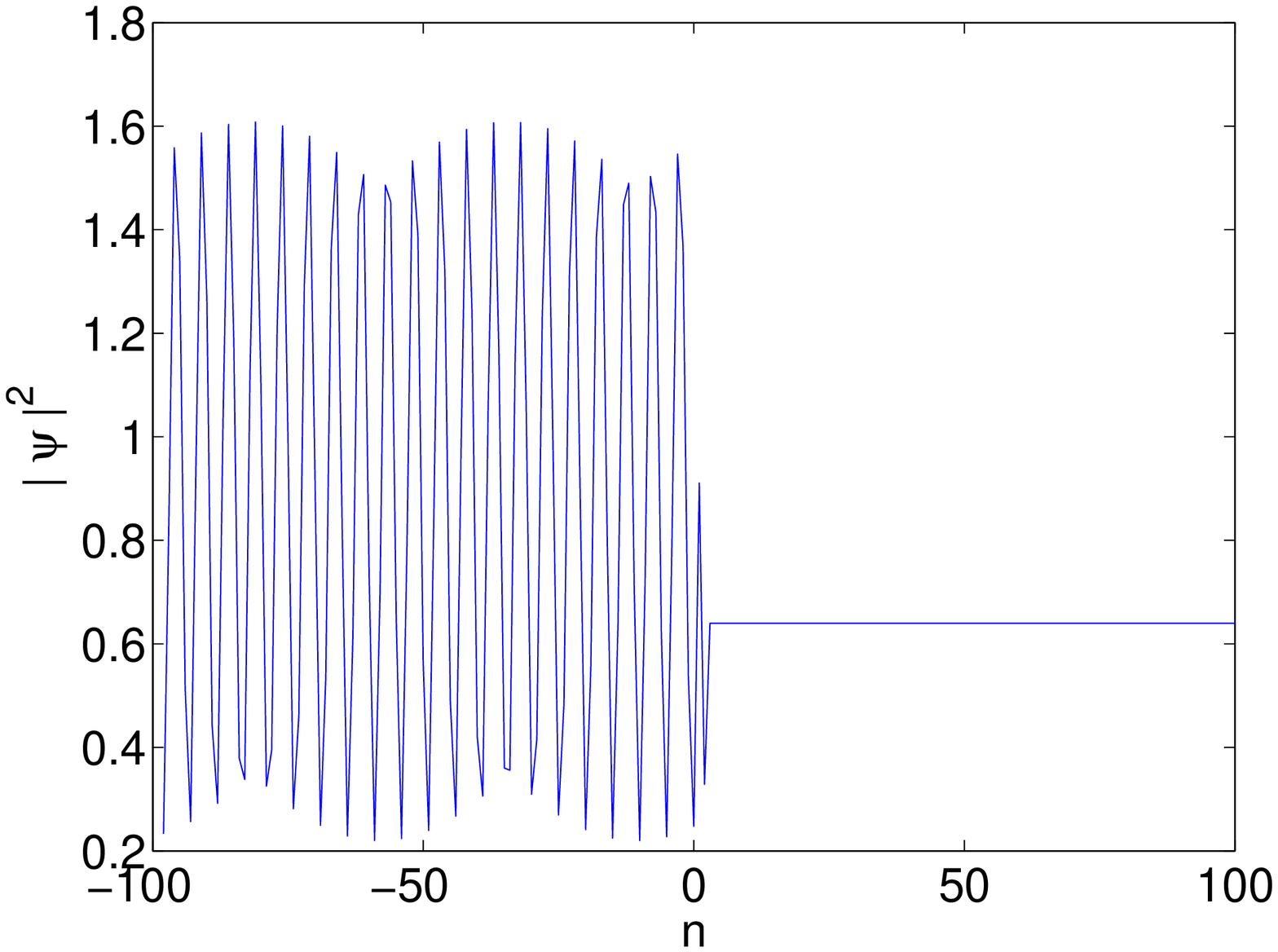}
\includegraphics[width=8cm,angle=0,clip]{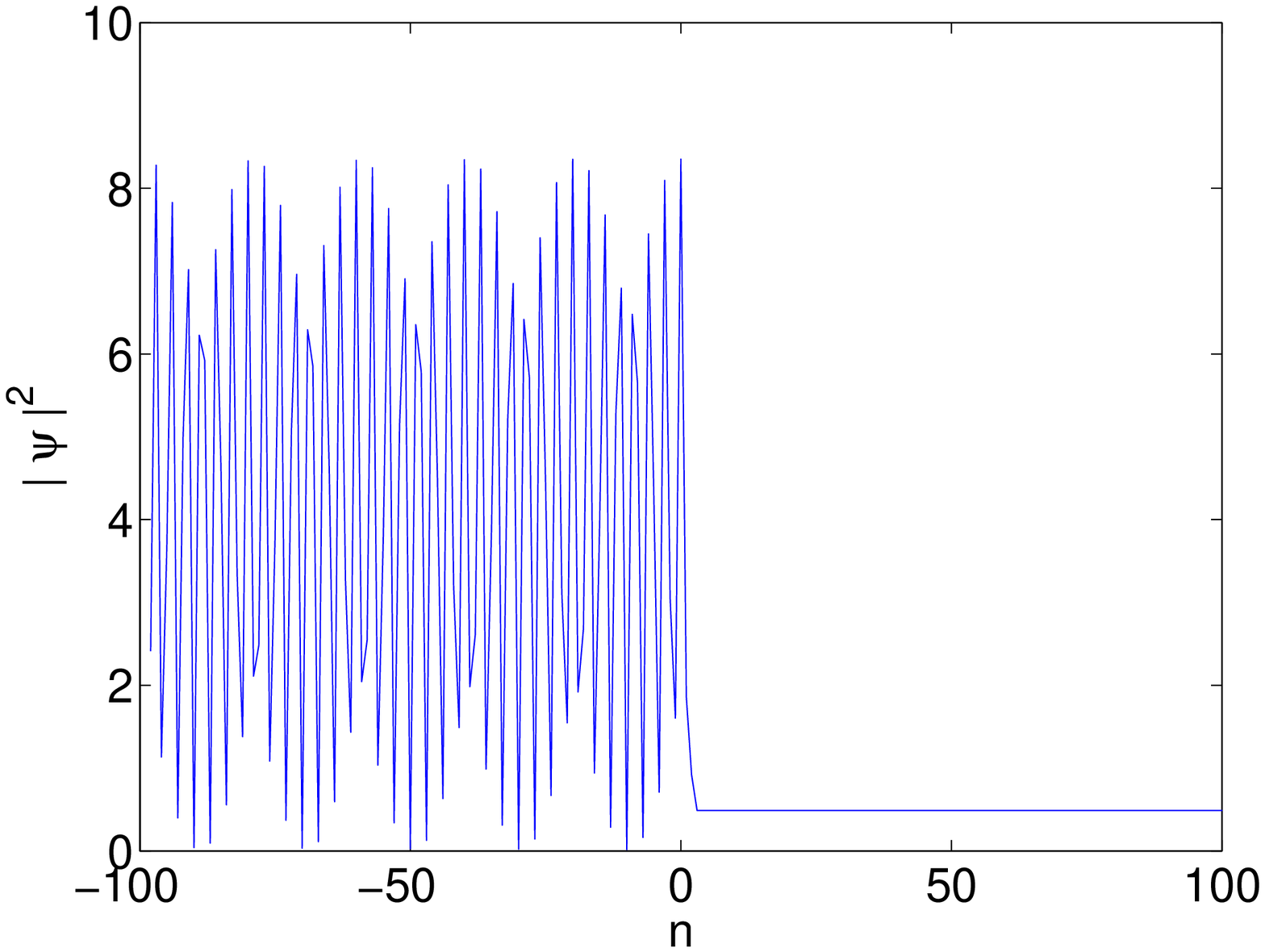}\\
\includegraphics[width=8cm,angle=0,clip]{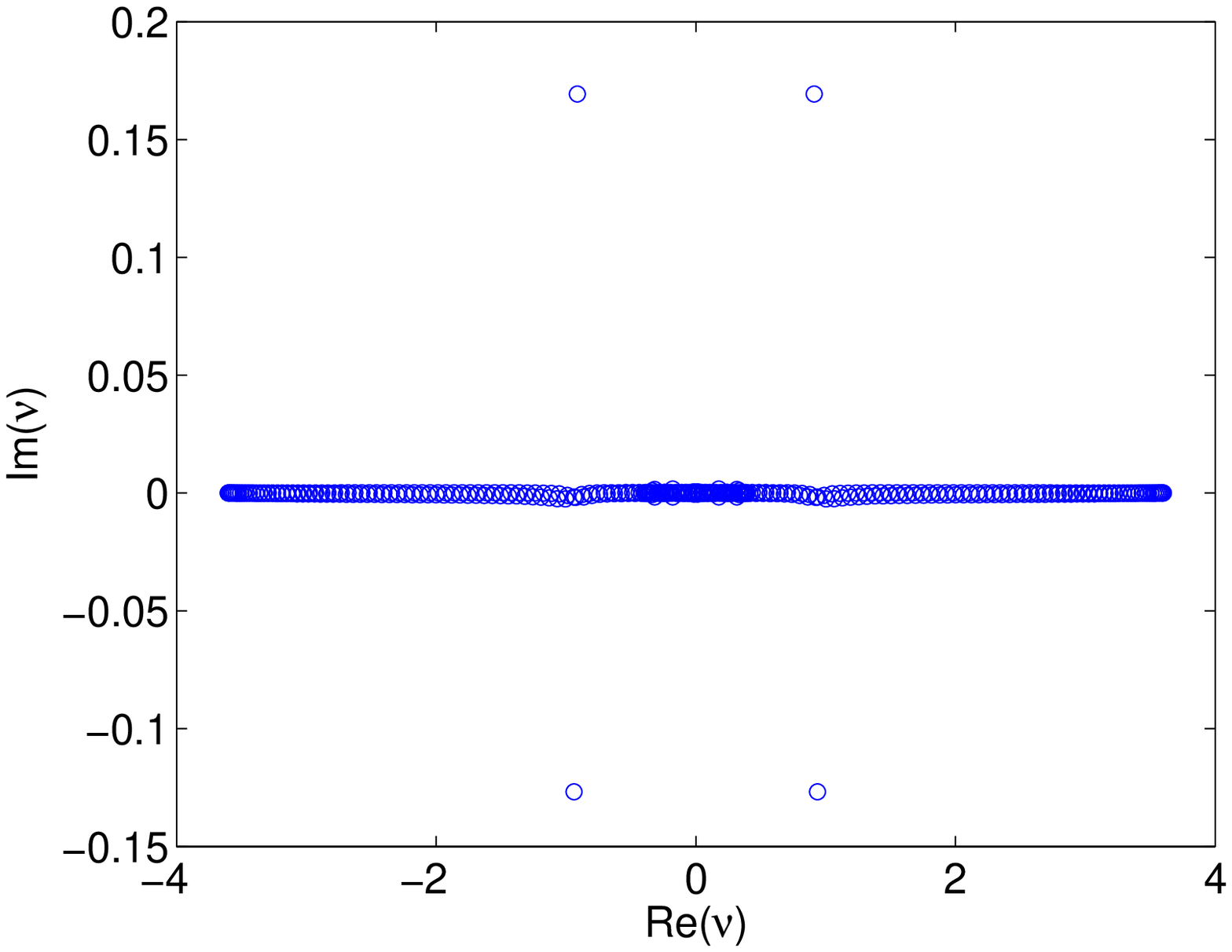}
\includegraphics[width=8cm,angle=0,clip]{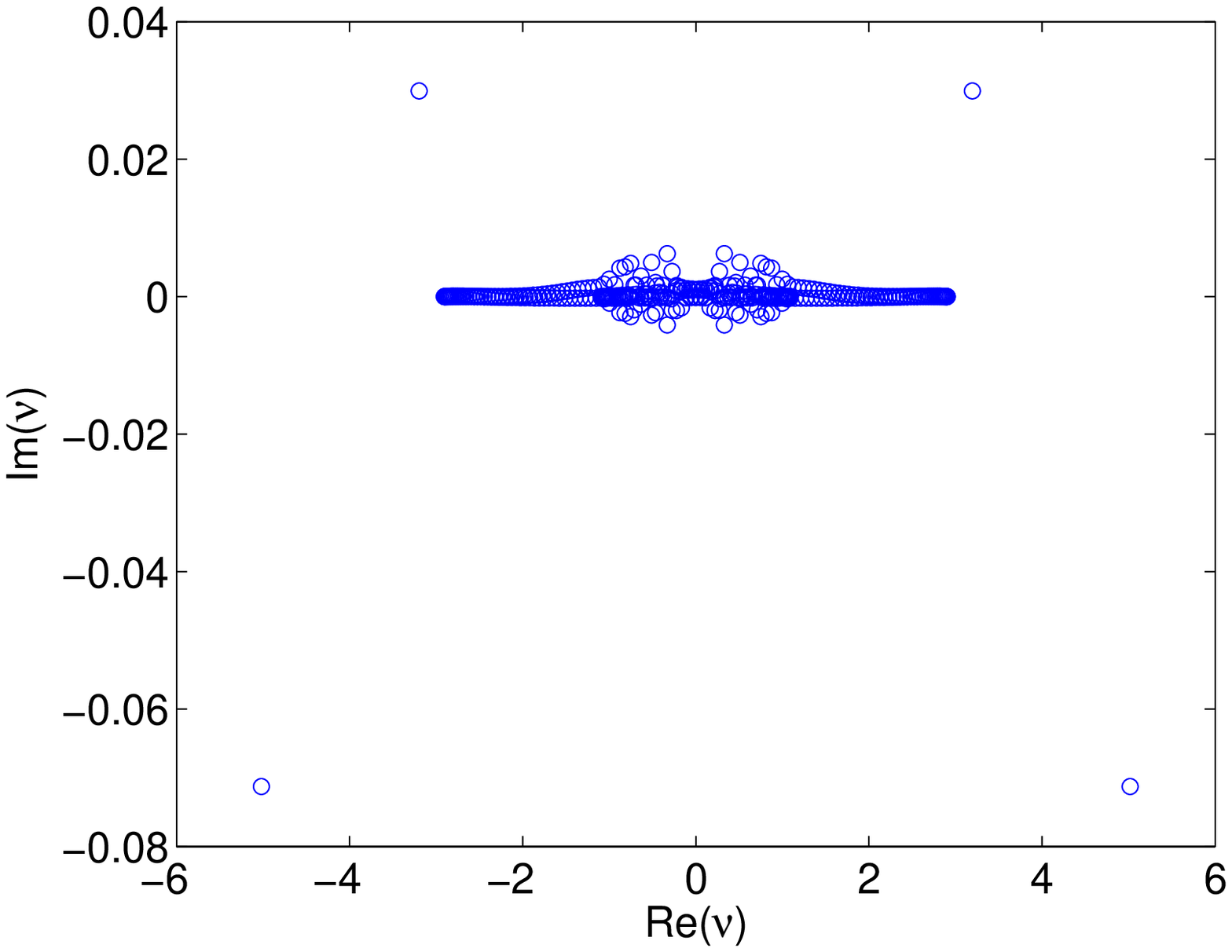}
\caption{The bottom panels show eigenvalues of the linearization of the example
solutions shown in the top panel.  
%and $|\phi_n(t)|$ (right).\newline
In  the PT-symmetric dimer (left), the parameters are 
$N=2$ with $k_0=2.5$, $T=0.8$, $\gamma=0.1$, while in the PT-symmetric
trimer (right), they
are: $N=3$ with $k_0=1.1$, $T=0.7$, $\gamma=0.1$.}
\label{figEigSpctm}
\end{center}
\end{figure}

\begin{figure}[tbp]
\begin{center}
\includegraphics[width=8cm,angle=0,clip]{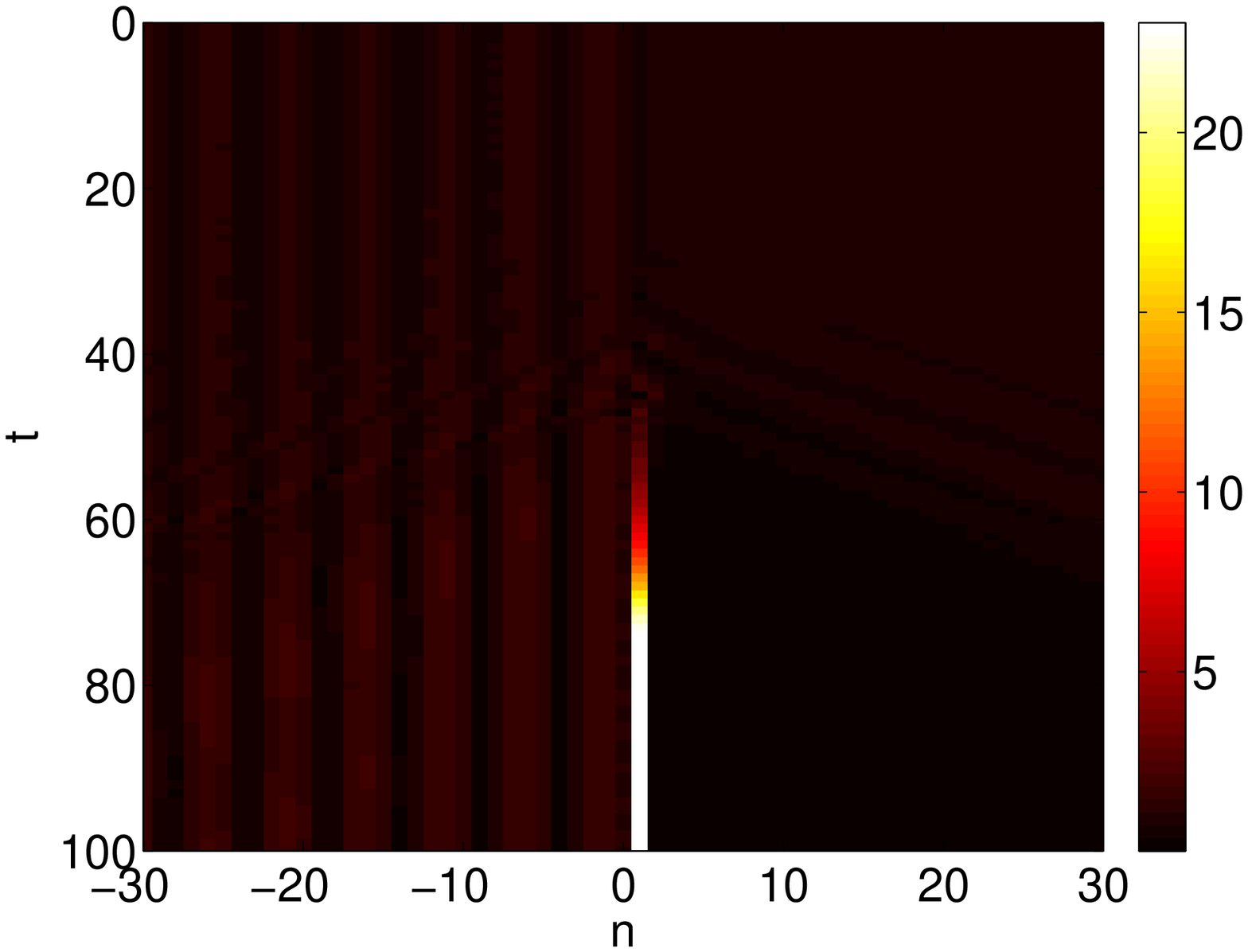}
\includegraphics[width=8cm,angle=0,clip]{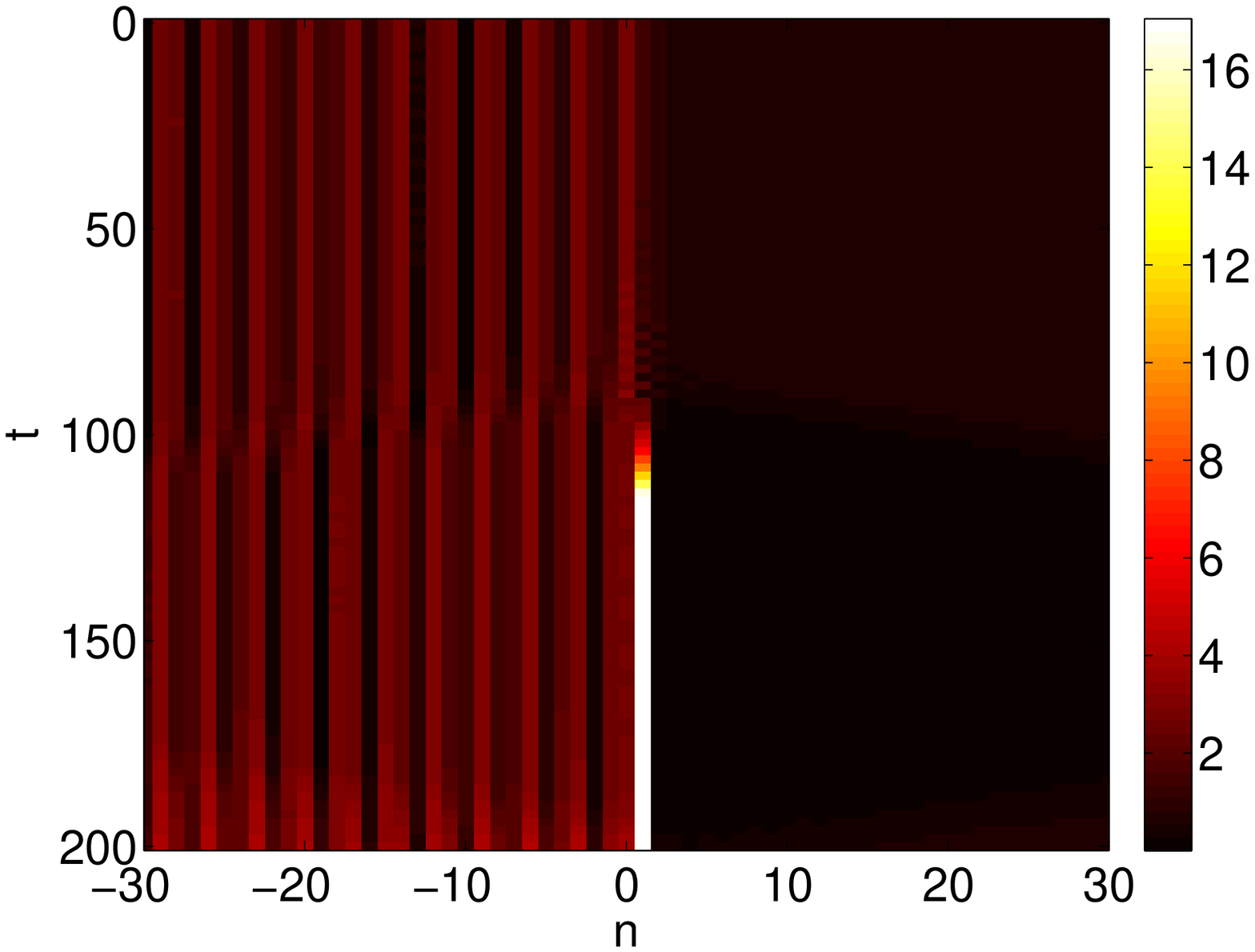}\\
\includegraphics[width=8cm,angle=0,clip]{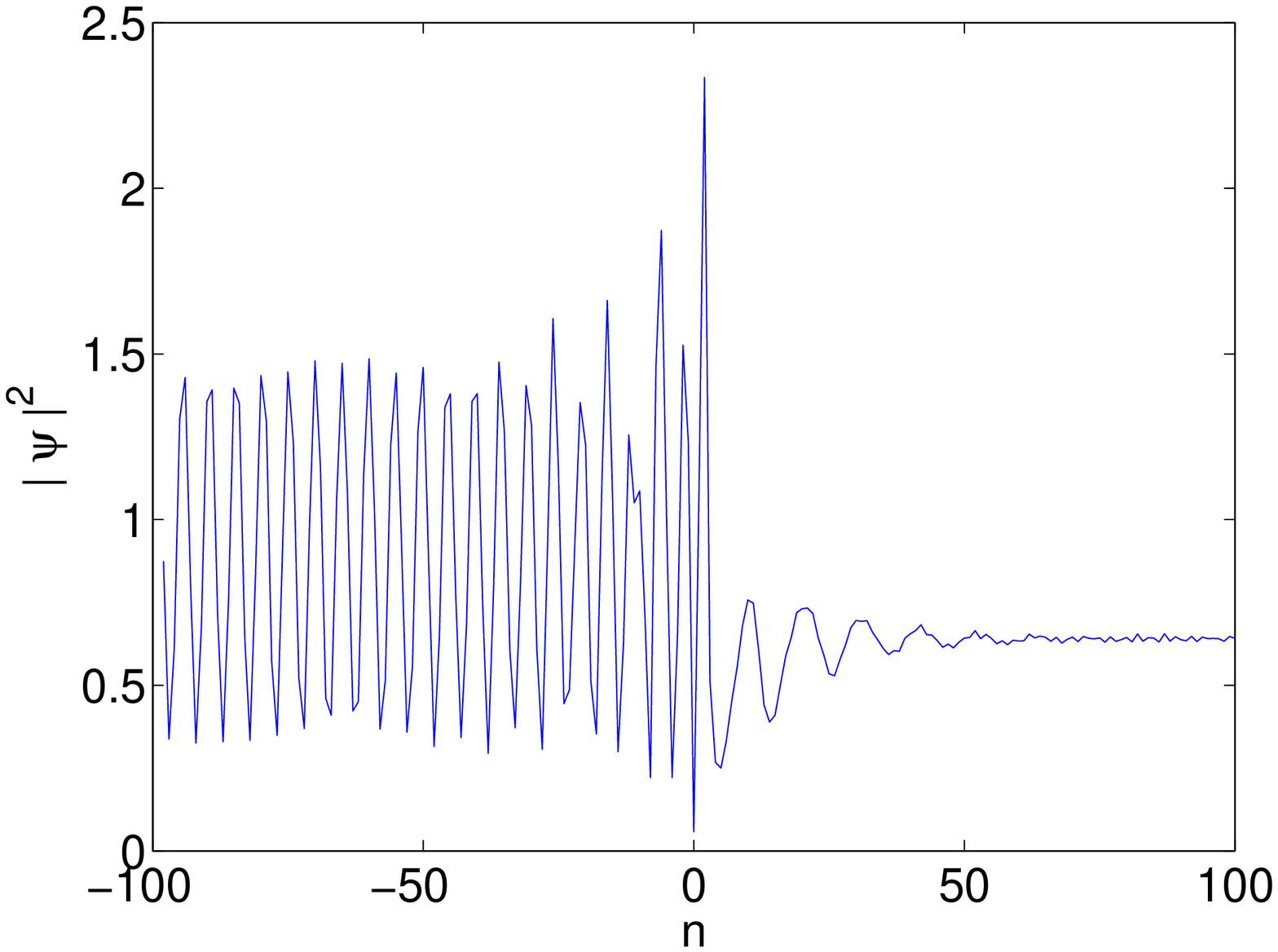}
\includegraphics[width=8cm,angle=0,clip]{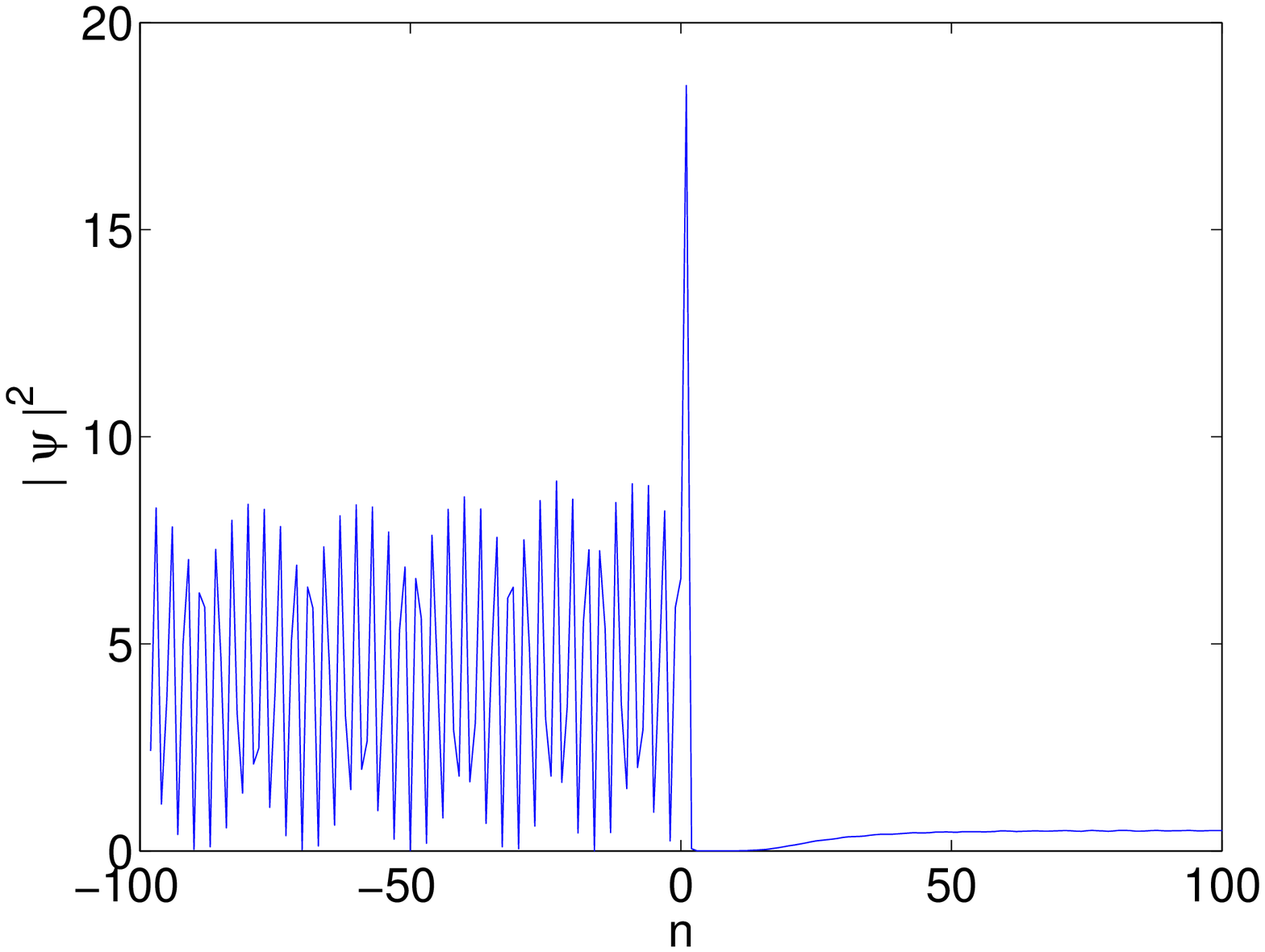}
\\
\includegraphics[width=8cm,angle=0,clip]{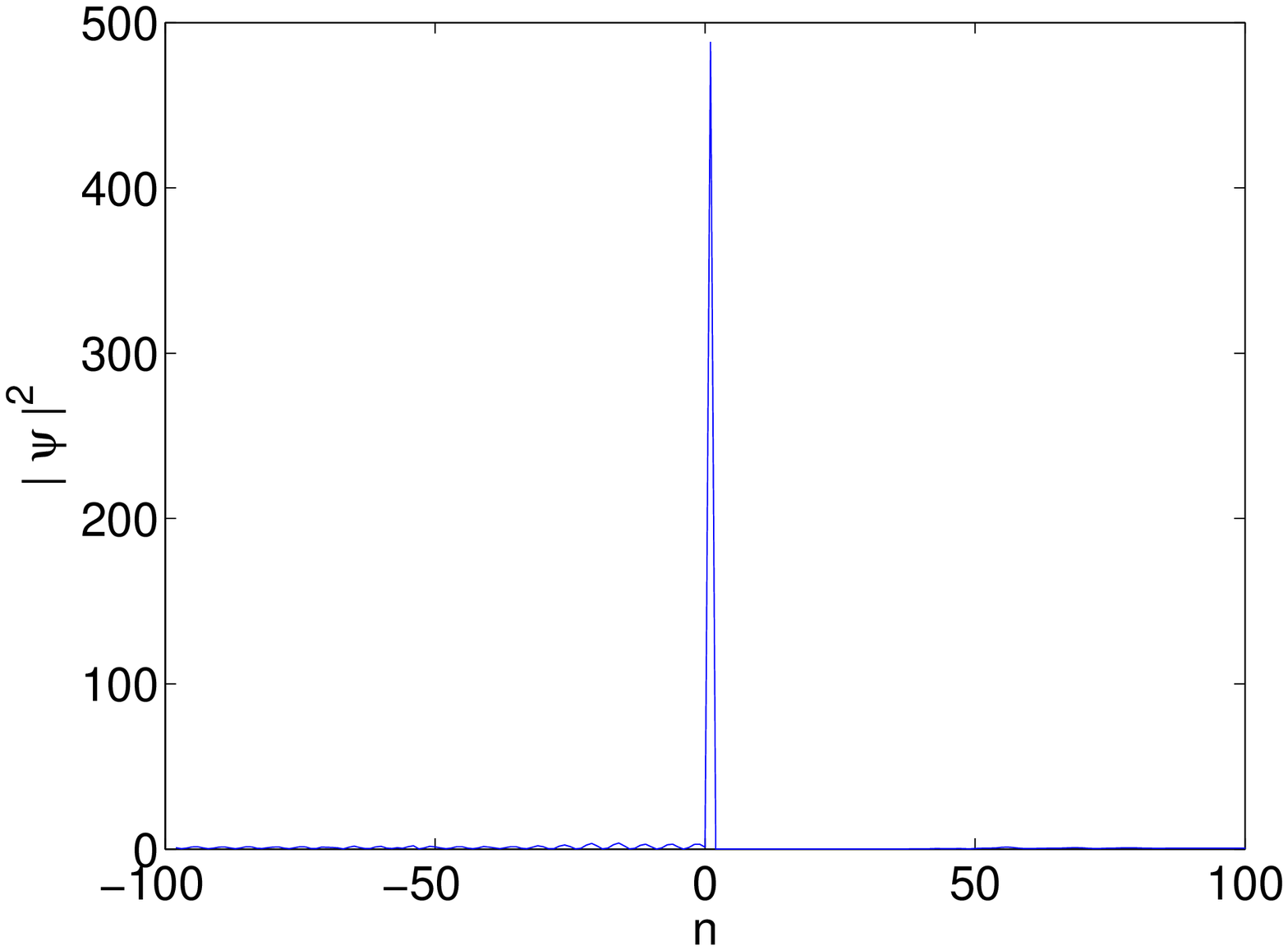}
\includegraphics[width=8cm,angle=0,clip]{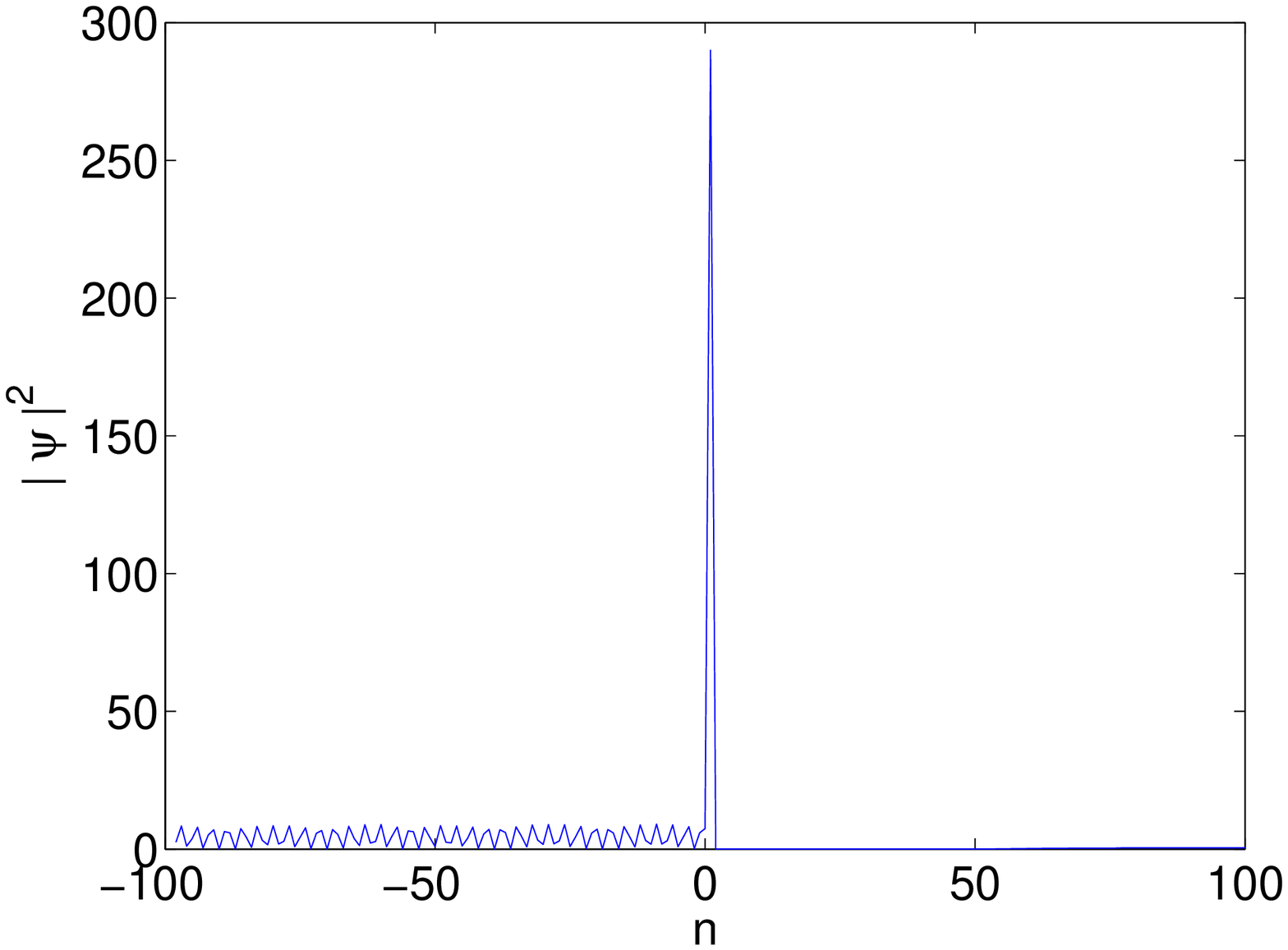}
\caption{The top panel shows the contour plot of the space $n$ - time
$t$ evolution of $|\phi_n|$ for the solutions/parameters 
corresponding to the left and right, respectively, panels of 
Fig.~\ref{figEigSpctm}.
%Profile solutions $|\phi_n|^2$
%\newlinec
The rest of the panels show individual snapshots of the solution
at $t=50, 80$ (for the left panels of the dimer case)
%\newline
and at $t=100, 120$ (for the right panels of the trimer case).}
\label{figProfiles}
\end{center}
\end{figure}

\begin{figure}[tbp]
\begin{center}
\includegraphics[width=8cm,angle=0,clip]{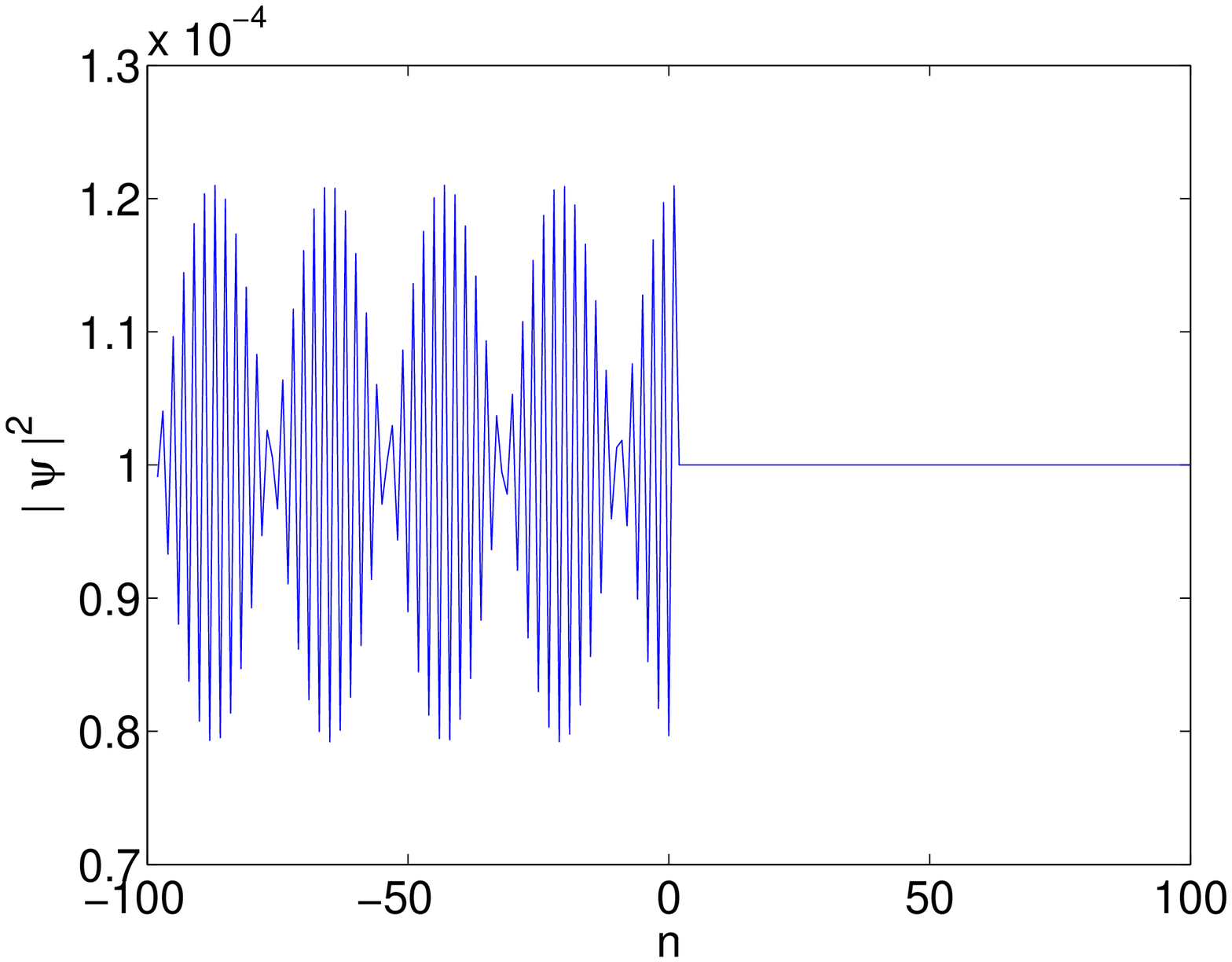}
\includegraphics[width=8cm,angle=0,clip]{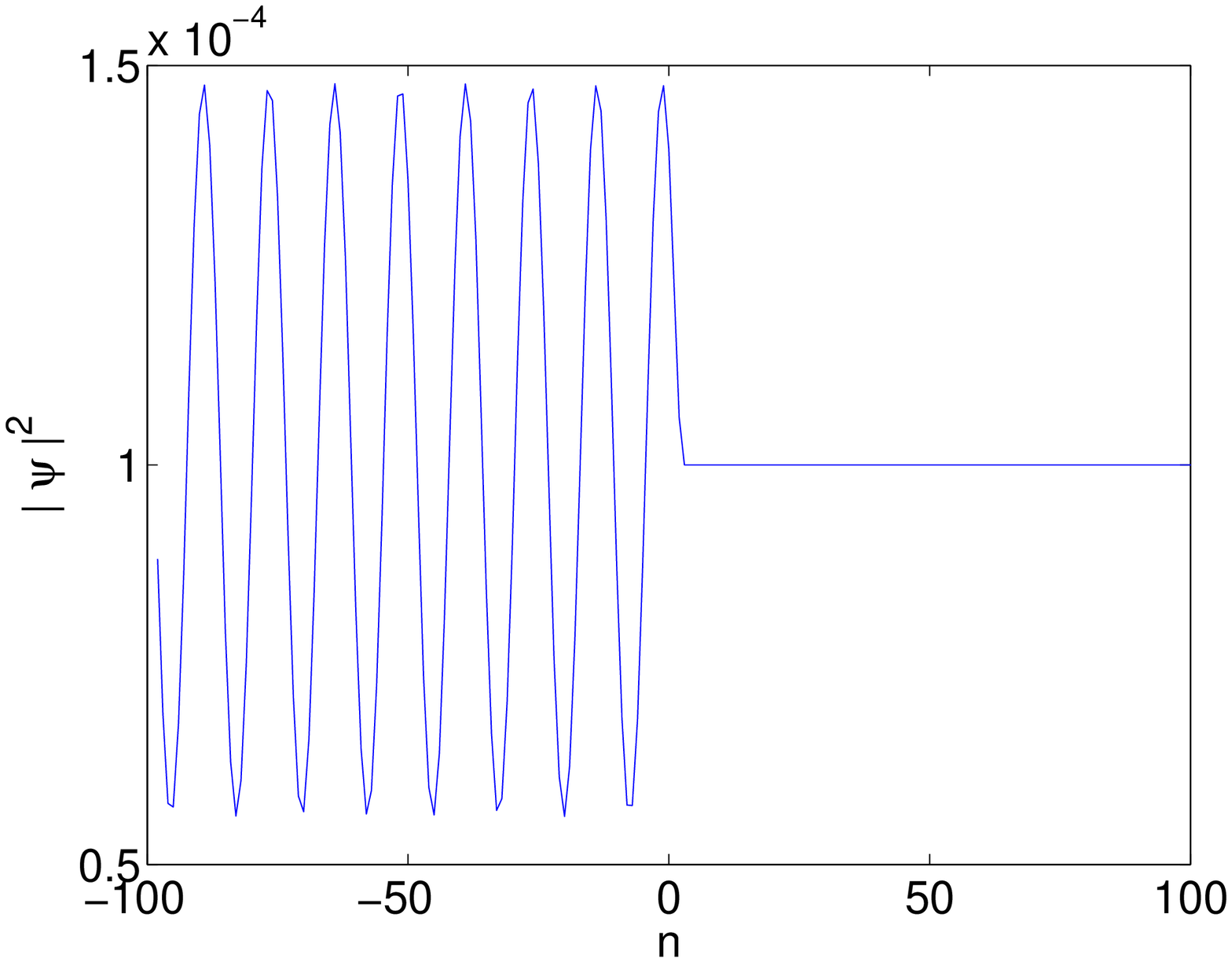}\\
\includegraphics[width=8cm,angle=0,clip]{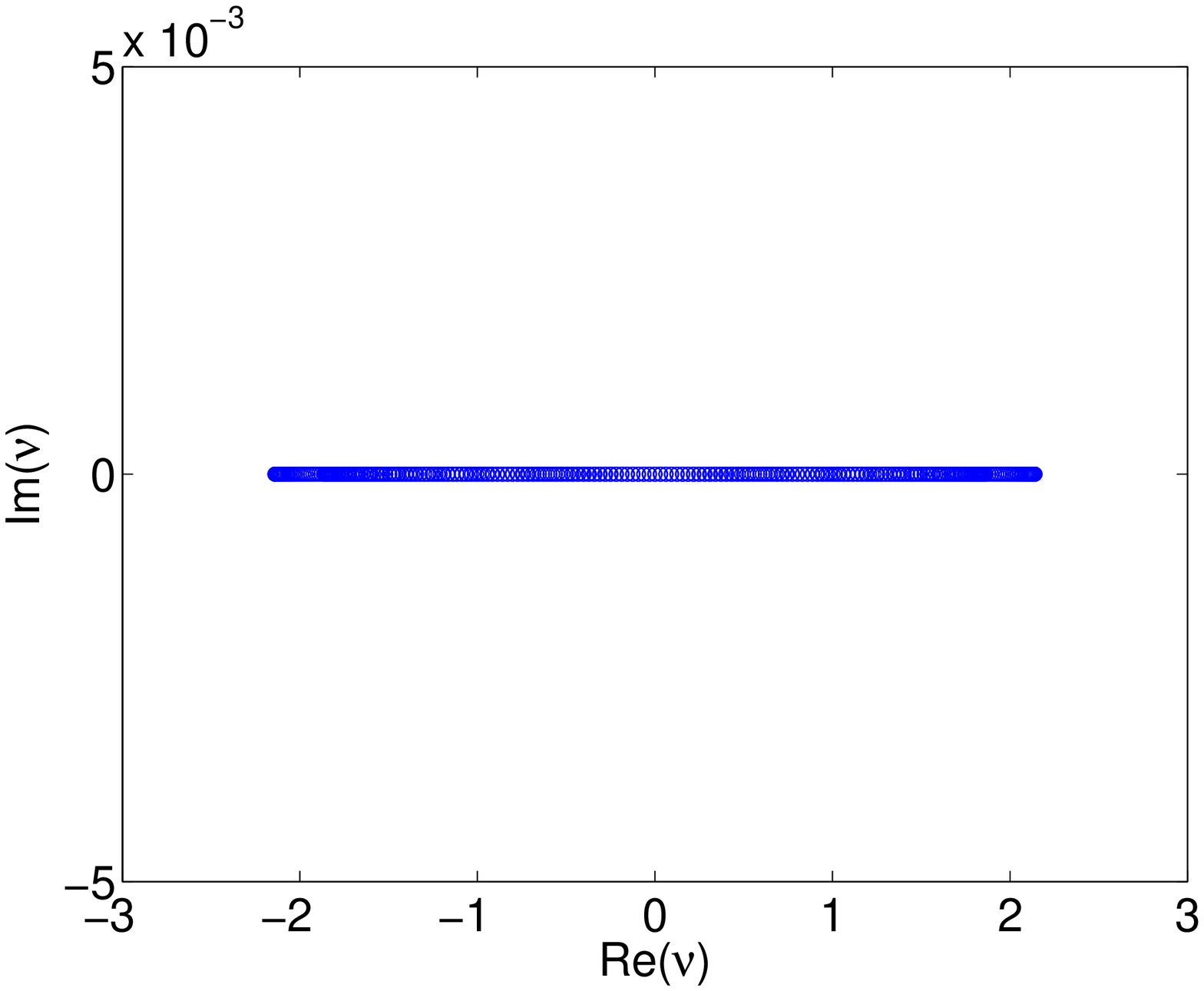}
\includegraphics[width=8cm,angle=0,clip]{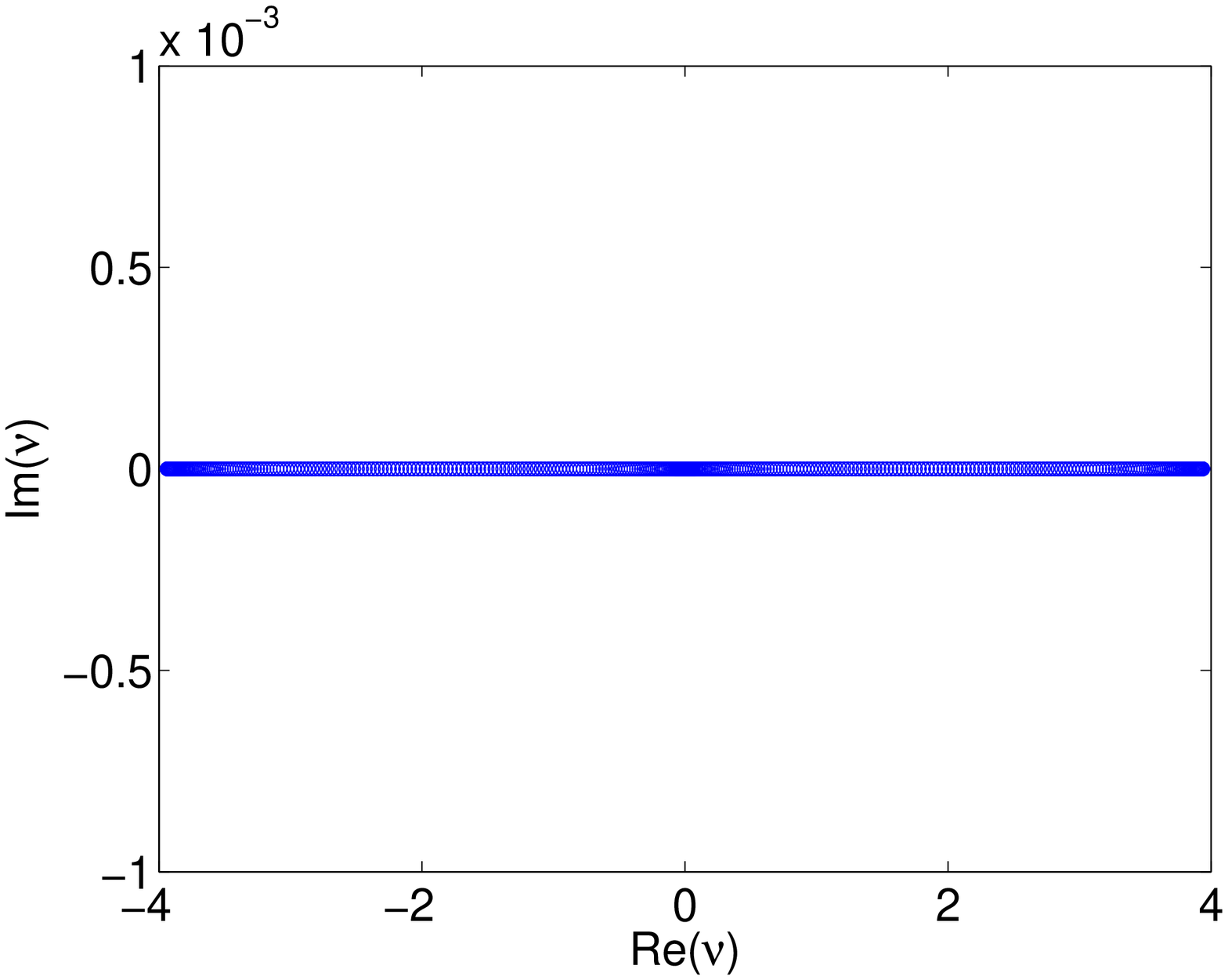}\\
\includegraphics[width=8cm,angle=0,clip]{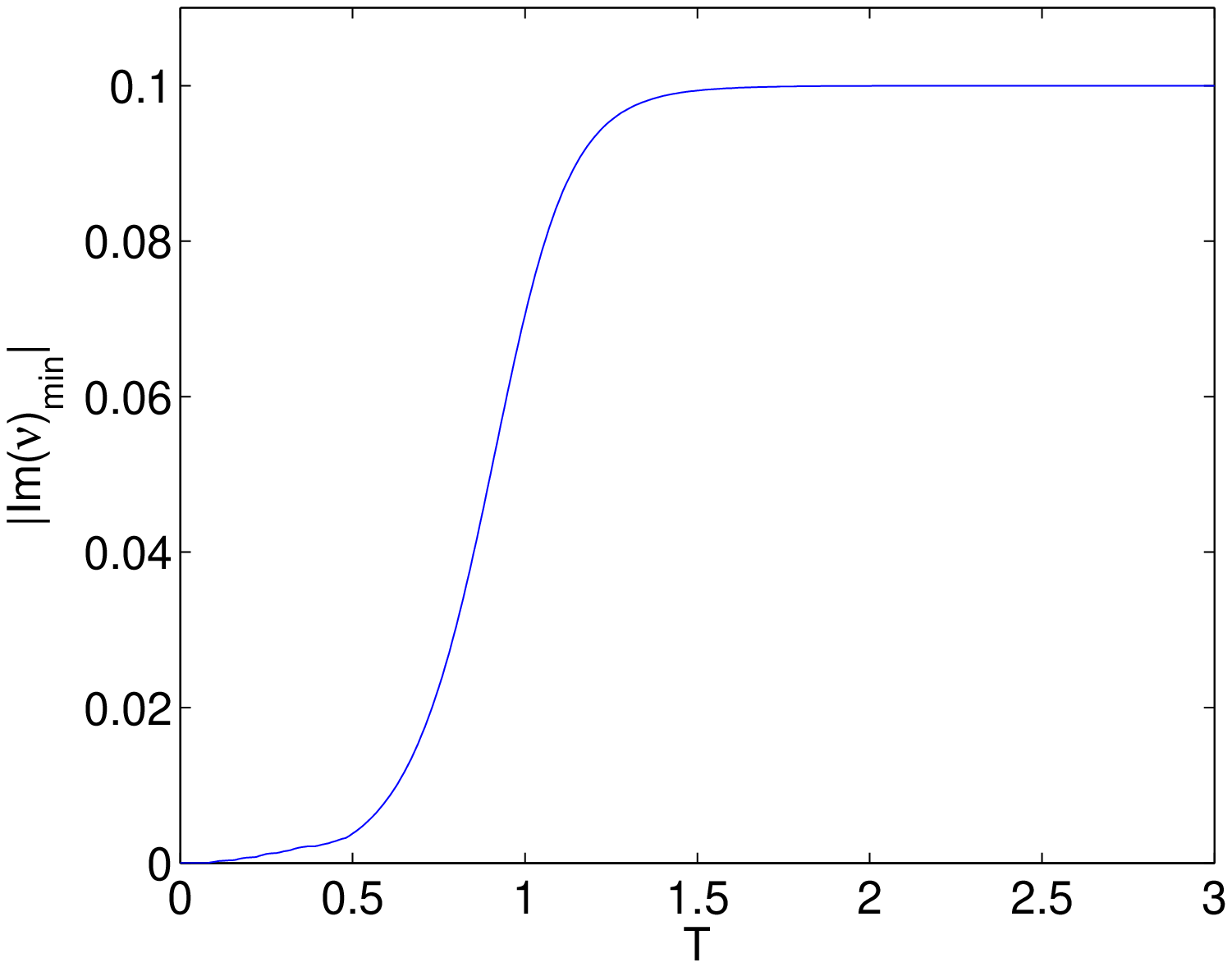}
%DimerMaxImEigk1_5gamma_1.eps}
\includegraphics[width=8cm,angle=0,clip]{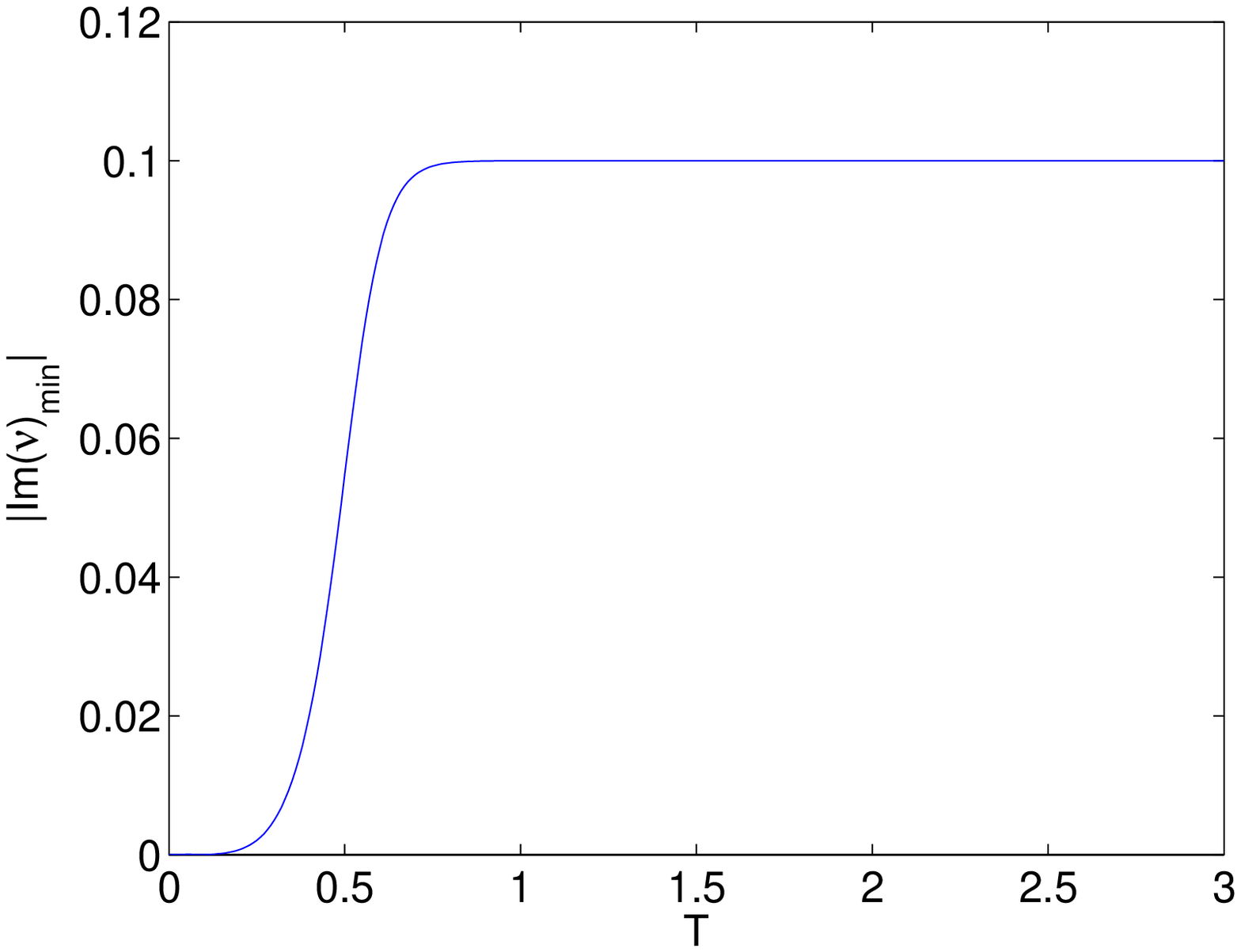}\\
%TrimerMaxImEigk0_25gamma_1.eps}\\
%\includegraphics[width=8cm,angle=0,clip]{N=3g1=_1T=_7k0=1_1spacetime.eps}
\caption{The top panels show the dimer and trimer solution
profiles for a small value of $T=0.01$ (and $k_0=1.5$, $\gamma=0.1$ for
the dimer, while $k_0=0.25$, $\gamma=0.1$ for the trimer). The middle
panels show the corresponding linear stability eigenfrequencies for this
near-linear, stable case. Finally, the bottom panels show
the dependence of the maximal growth rate (maximal negative imaginary
eigenfrequency) as a function of the transmission parameter $T$ of the
wave, showing the significant instability enhancement upon 
progressive departure from the linear limit of $T \rightarrow 0$.}
\label{figEigSpctm_add}
\end{center}
\end{figure}

\begin{figure}[tbp]
\begin{center}
\includegraphics[width=8cm,angle=0,clip]{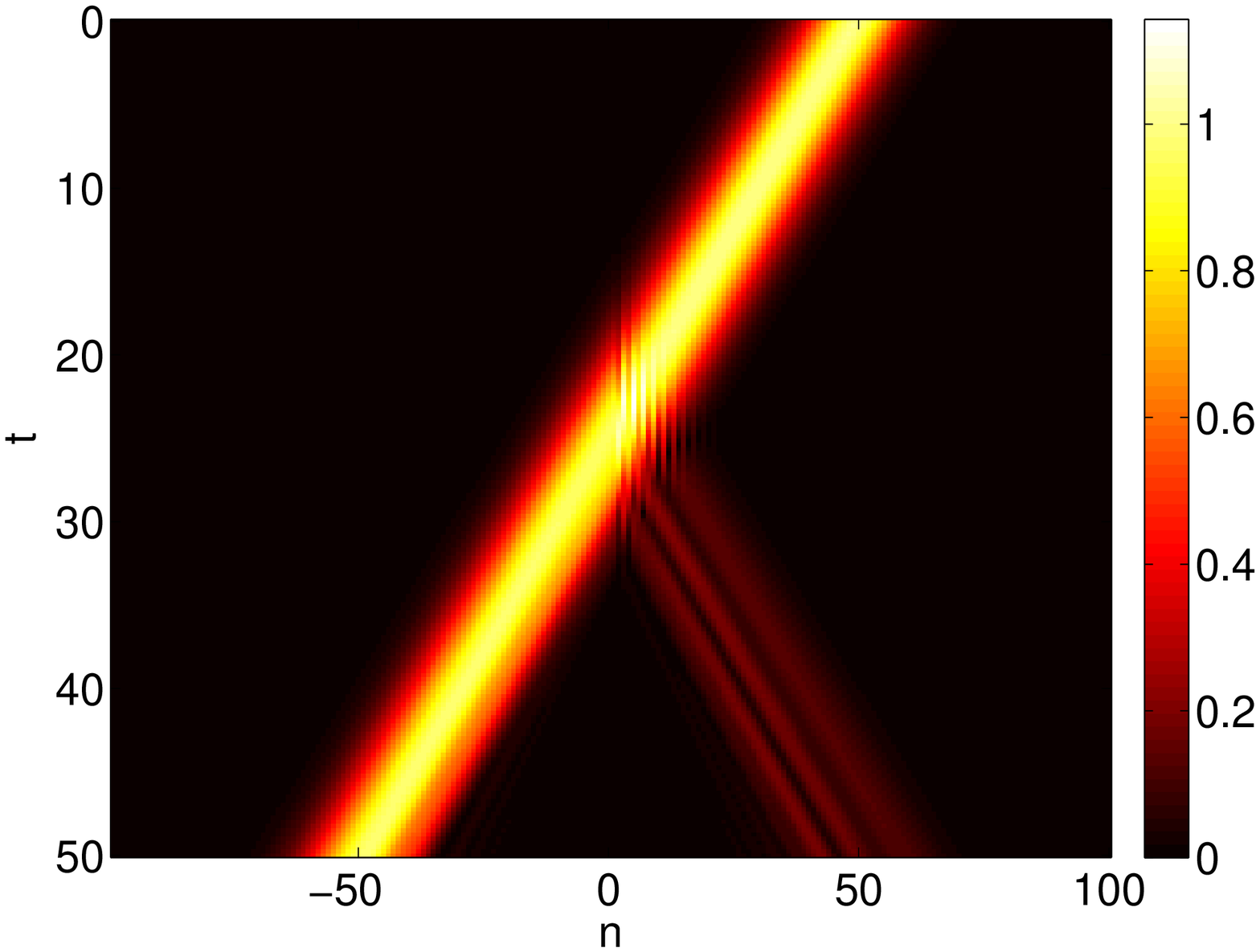}
\includegraphics[width=8cm,angle=0,clip]{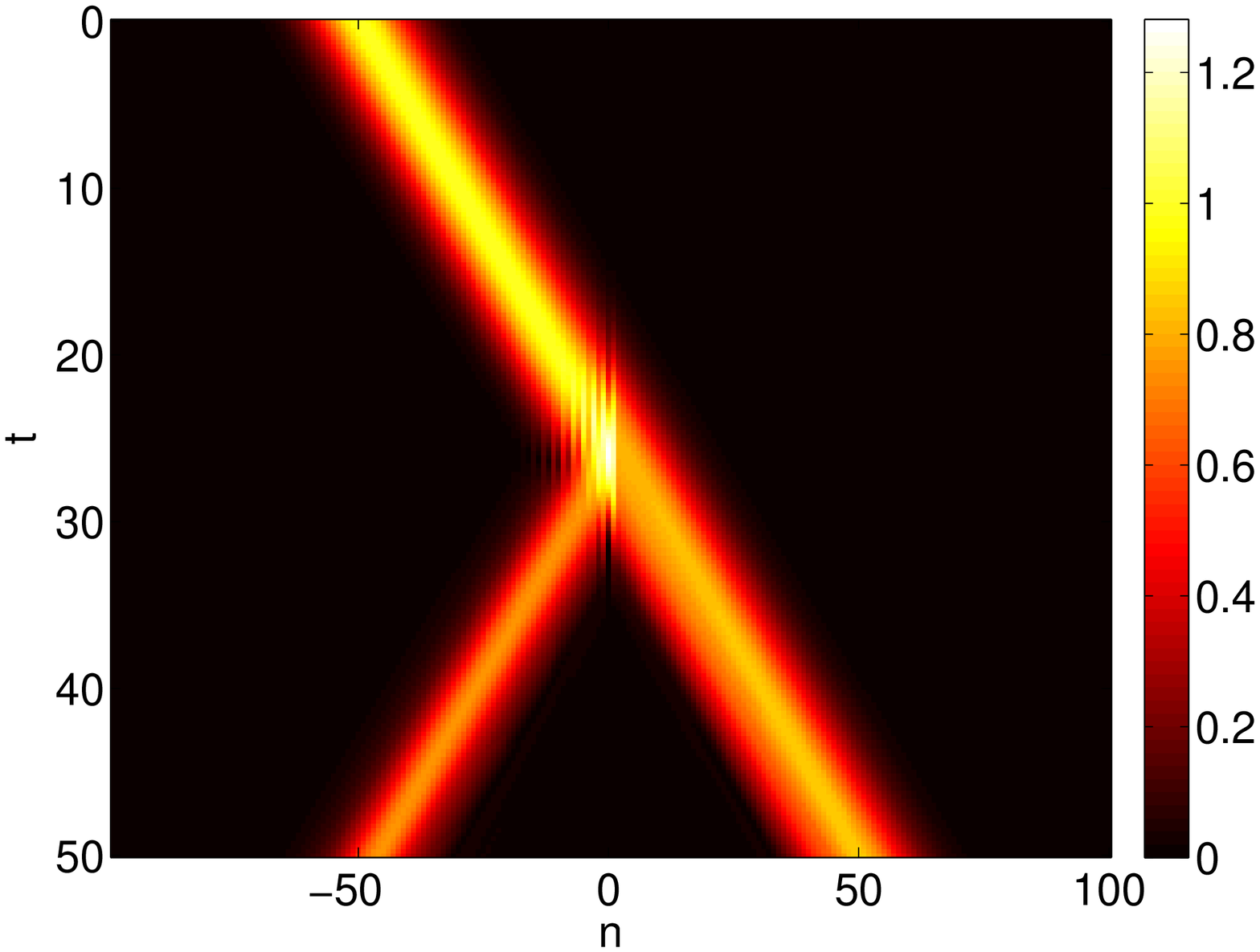}\\
\includegraphics[width=8cm,angle=0,clip]{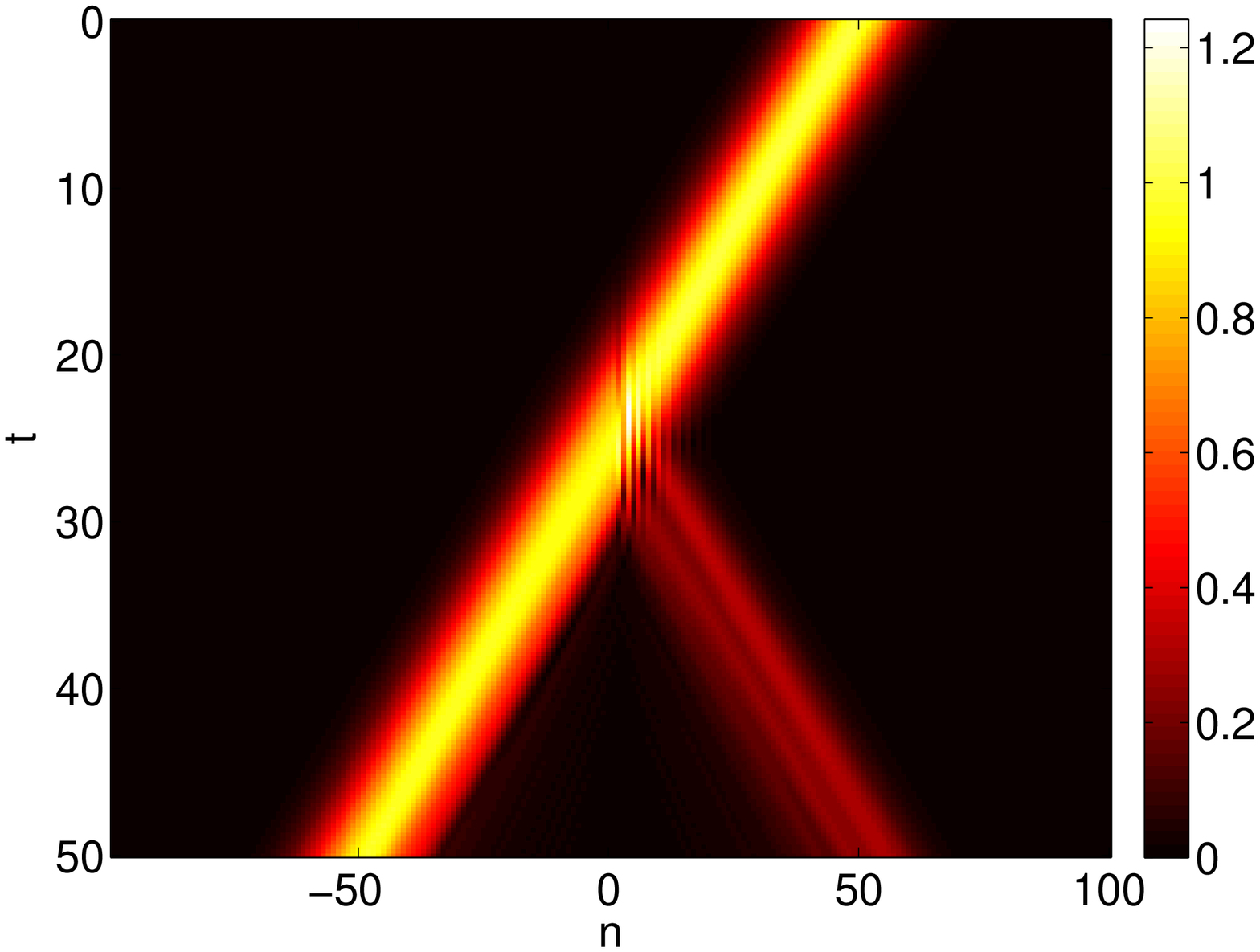}
\includegraphics[width=8cm,angle=0,clip]{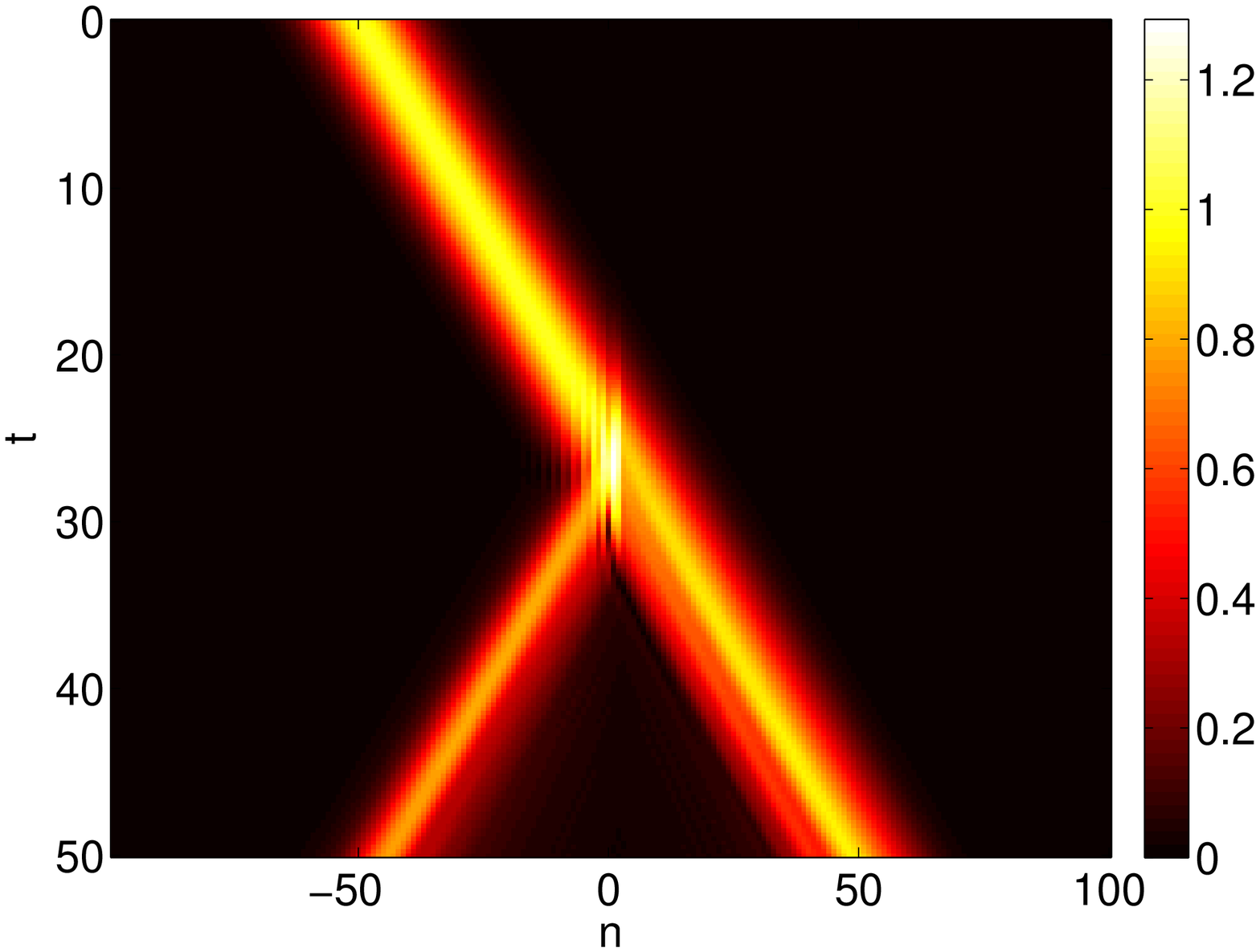}
\caption{The space-time contour plot of $|\phi_n(t)|^2$ is shown
for left and right incidence on the PT-symmetric dimer (top panels)
and on the PT-symmetric trimer (bottom panels).  
Top:  dimer, $N=2$ with $n_0=-50$, $s=10$, $k_0=-\pi/2$ (left) and $k_0=\pi/2$ (right);
Bottom:  trimer, $N=3$ with $n_0=-50$, $s=10$, $k_0=-\pi/2$ (left) and $k_0=\pi/2$ (right), 
}
\label{figGauss}
\end{center}
\end{figure}

\begin{figure}[tbp]
\begin{center}
\includegraphics[width=8cm,angle=0,clip]{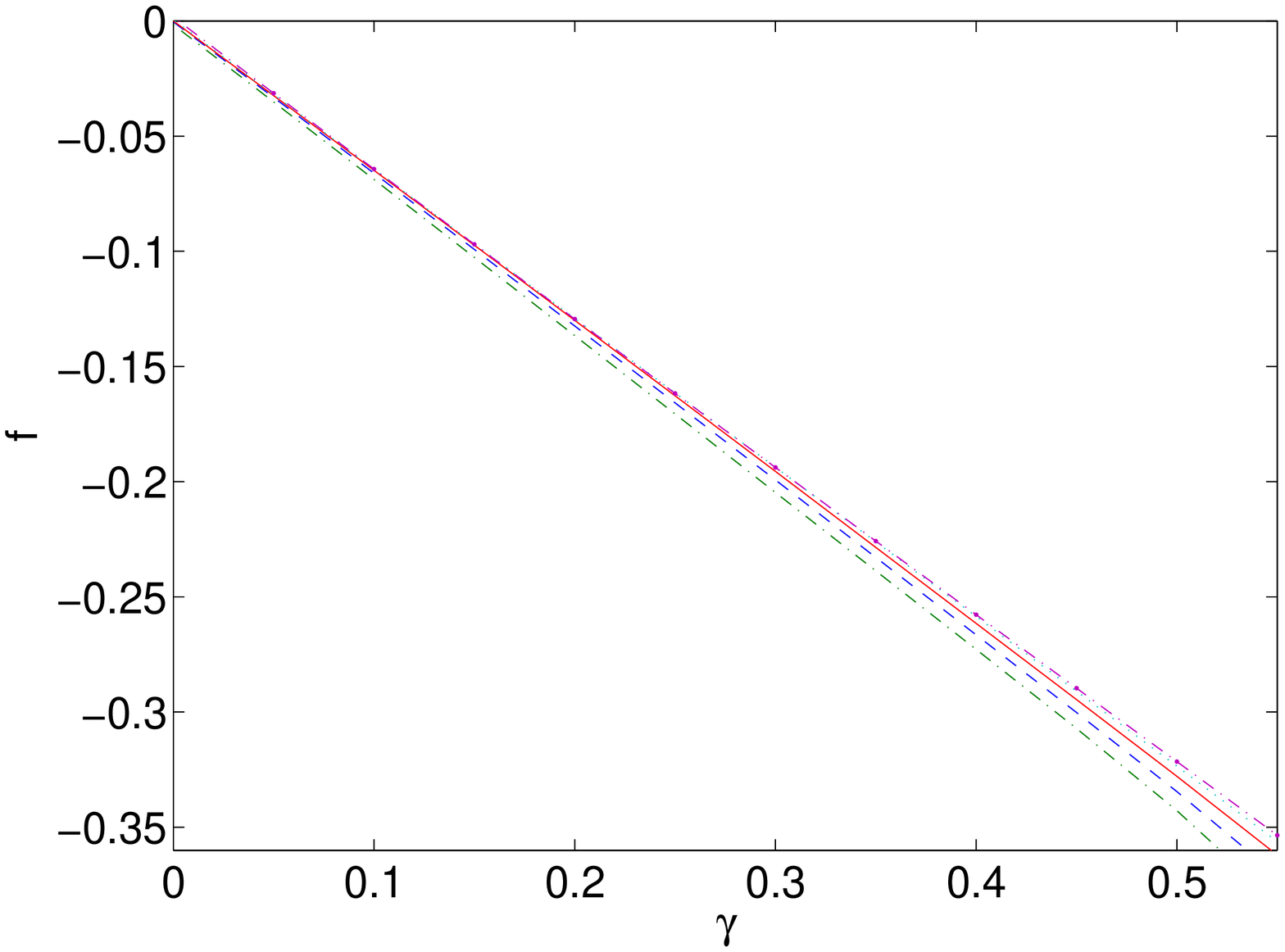}
\includegraphics[width=8cm,angle=0,clip]{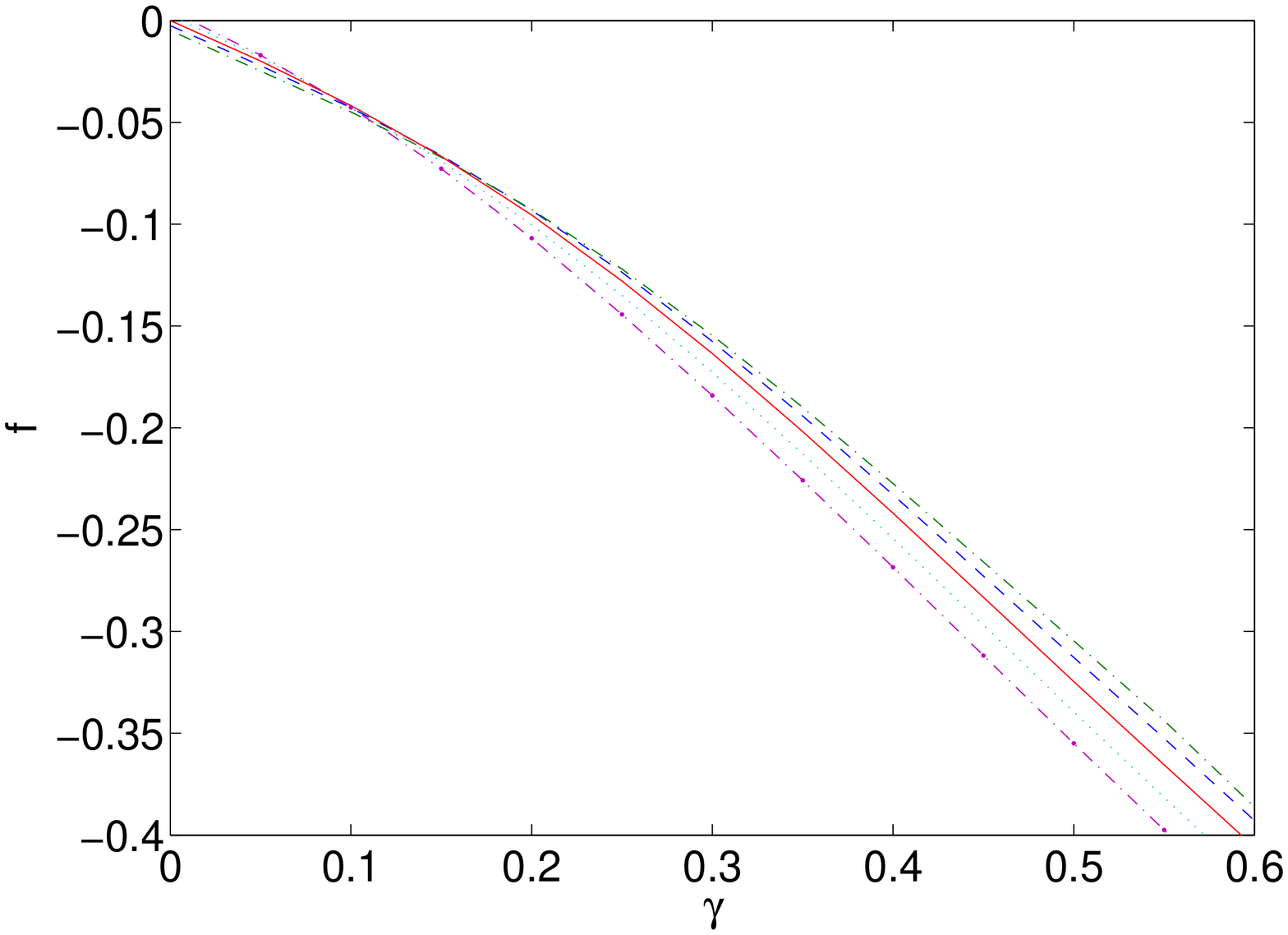}
\caption{Rectifying factor for the incidence of a Gaussian wavepacket
on the PT-symmetric dimer and trimer.
Left:  dimer, $N=2$ with { $\alpha_1=1-\Delta\alpha/2, \alpha_2=1+\Delta\alpha/2$ and} $k_0=\pi/2$, $n_0=-50$, $s=10$;
Right:  trimer, $N=3$ with { $\alpha_1=1-\Delta\alpha/2, \alpha_2=1, \alpha_3=1+\Delta\alpha/2$ and} $k_0=\pi/2$, $n_0=-50$, $s=10$.  { The following values of $\Delta\alpha$ are plotted:  $-0.5$ (dashed lines), 
$-0.25$ (dashed-dotted), $0$ (solid), $0.25$ (dotted), $0.5$ (dashed with bold dots).}
}
\label{figtdiff}
\end{center}
\end{figure}

\begin{figure}[tbp]
\begin{center}
\includegraphics[width=8cm,height=5.75cm,angle=0,clip]{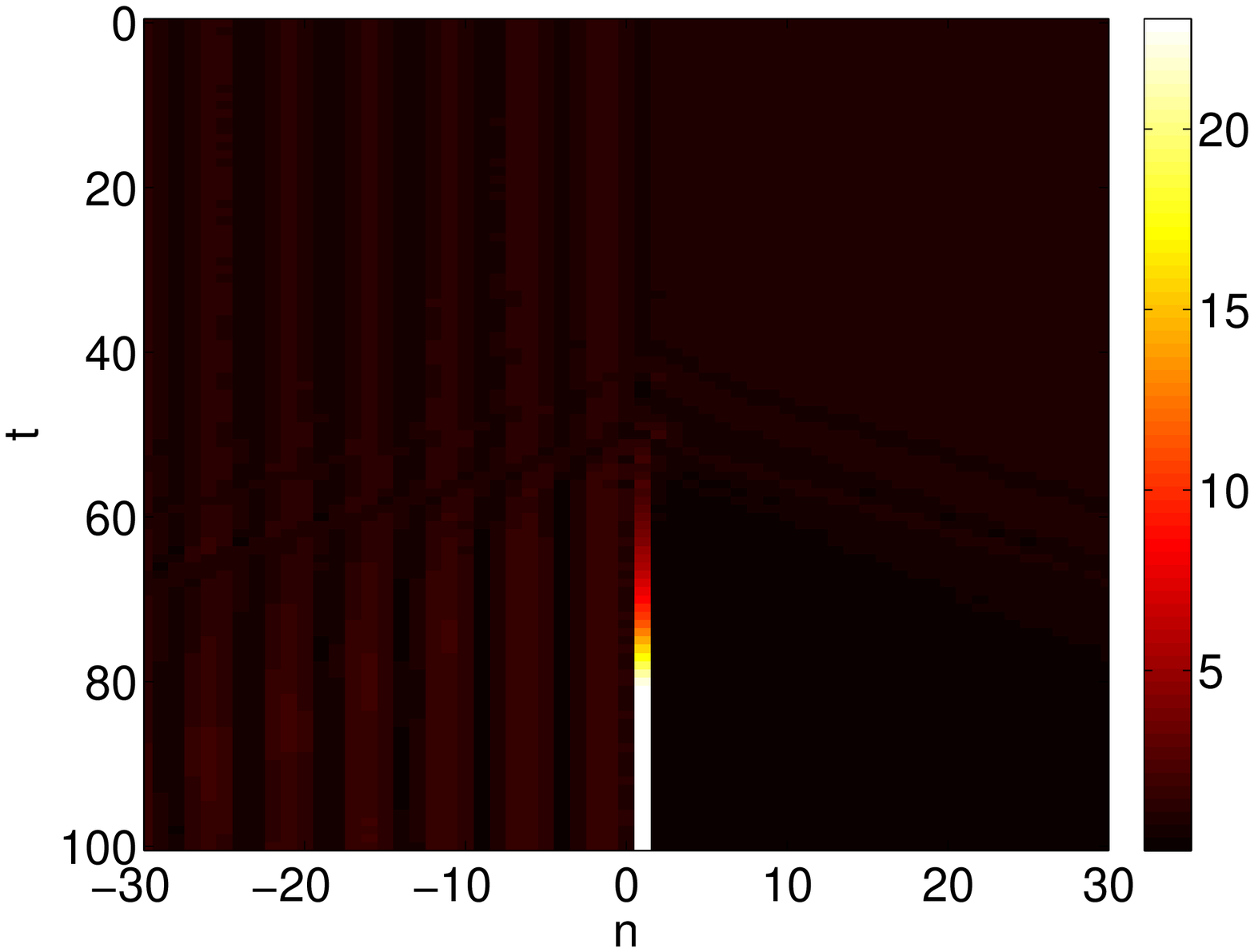}
\includegraphics[width=8cm,height=5.75cm,angle=0,clip]{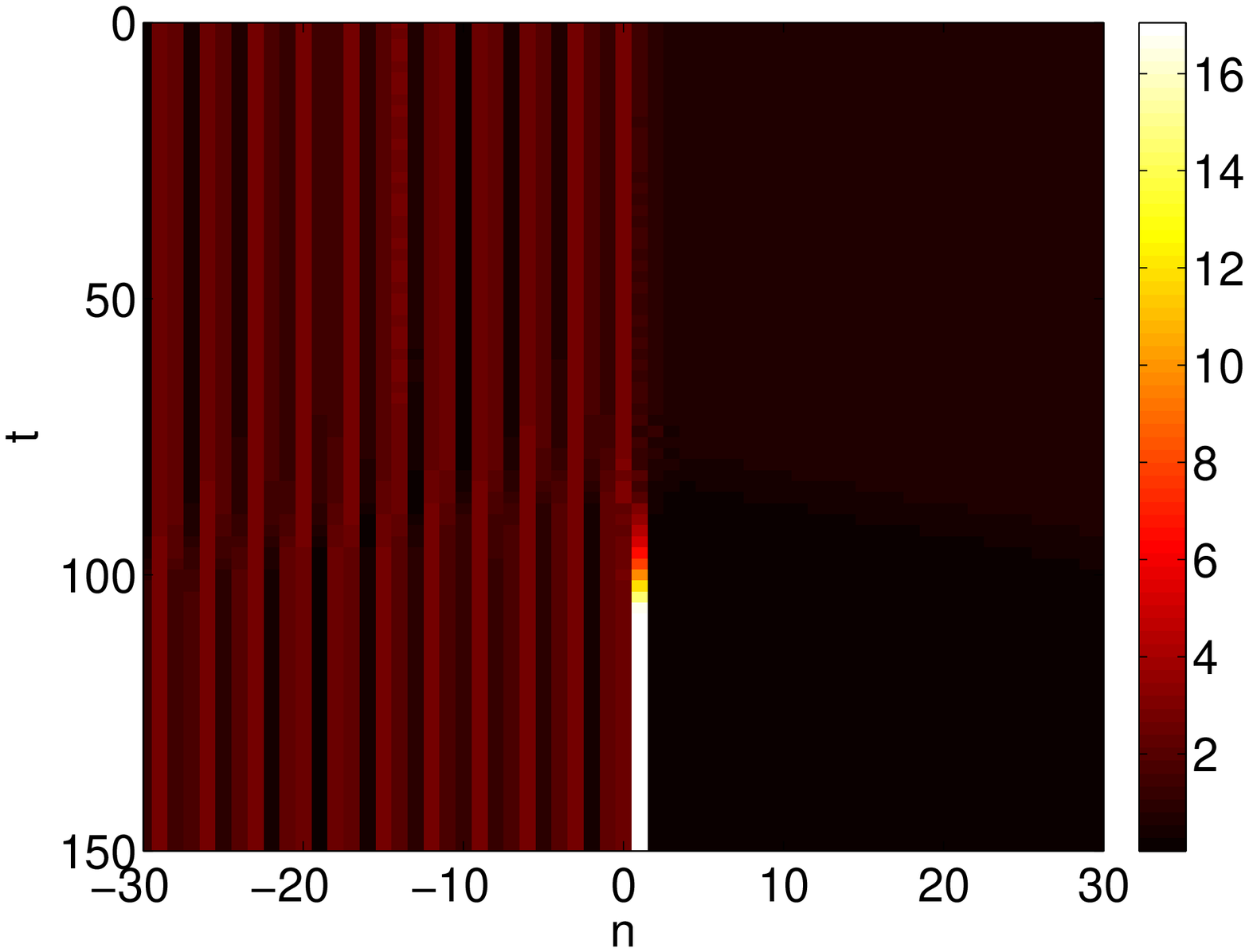}\\
\includegraphics[width=8cm,height=6.2cm,angle=0,clip]{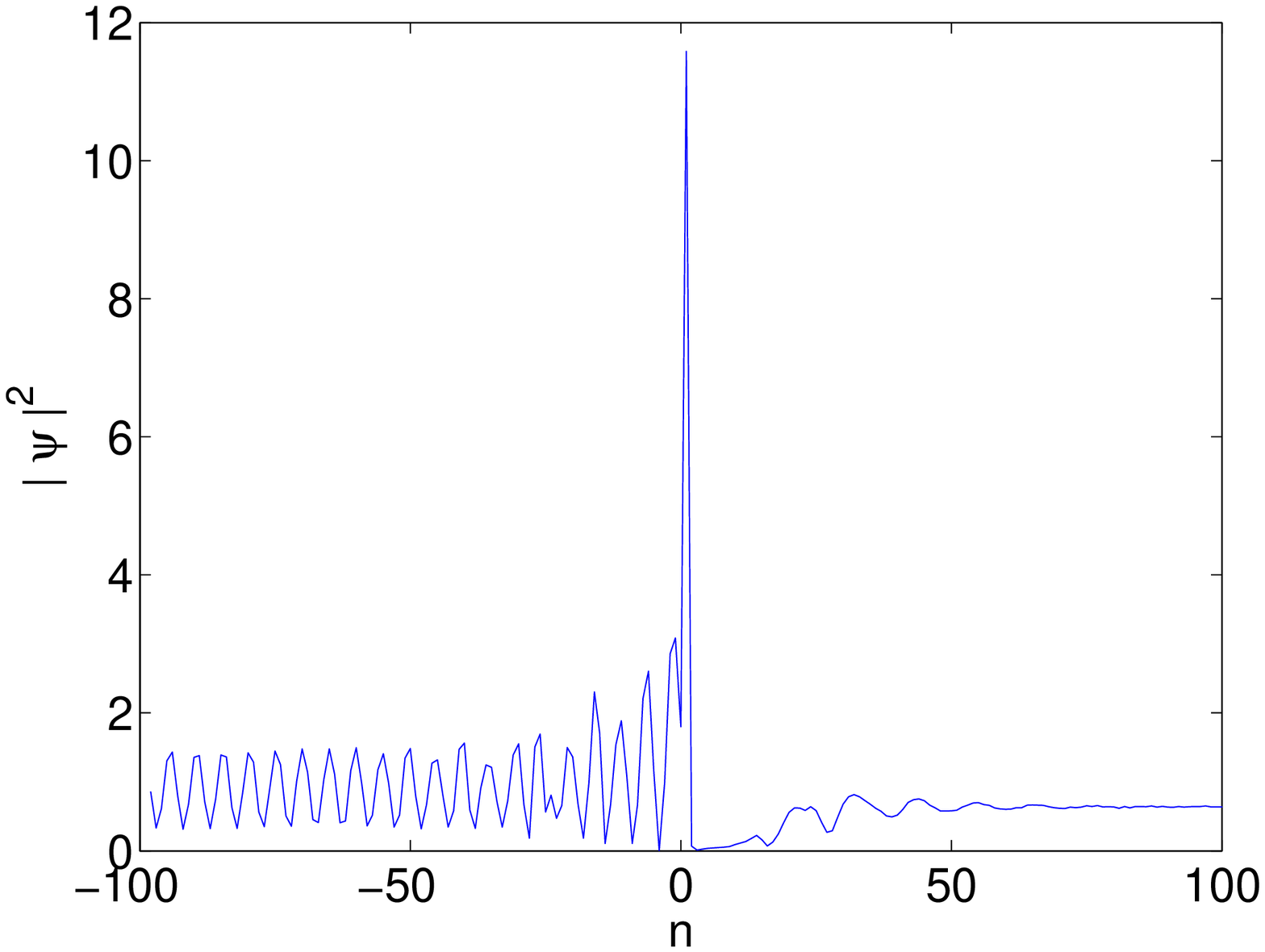}
\includegraphics[width=8cm,angle=0,clip]{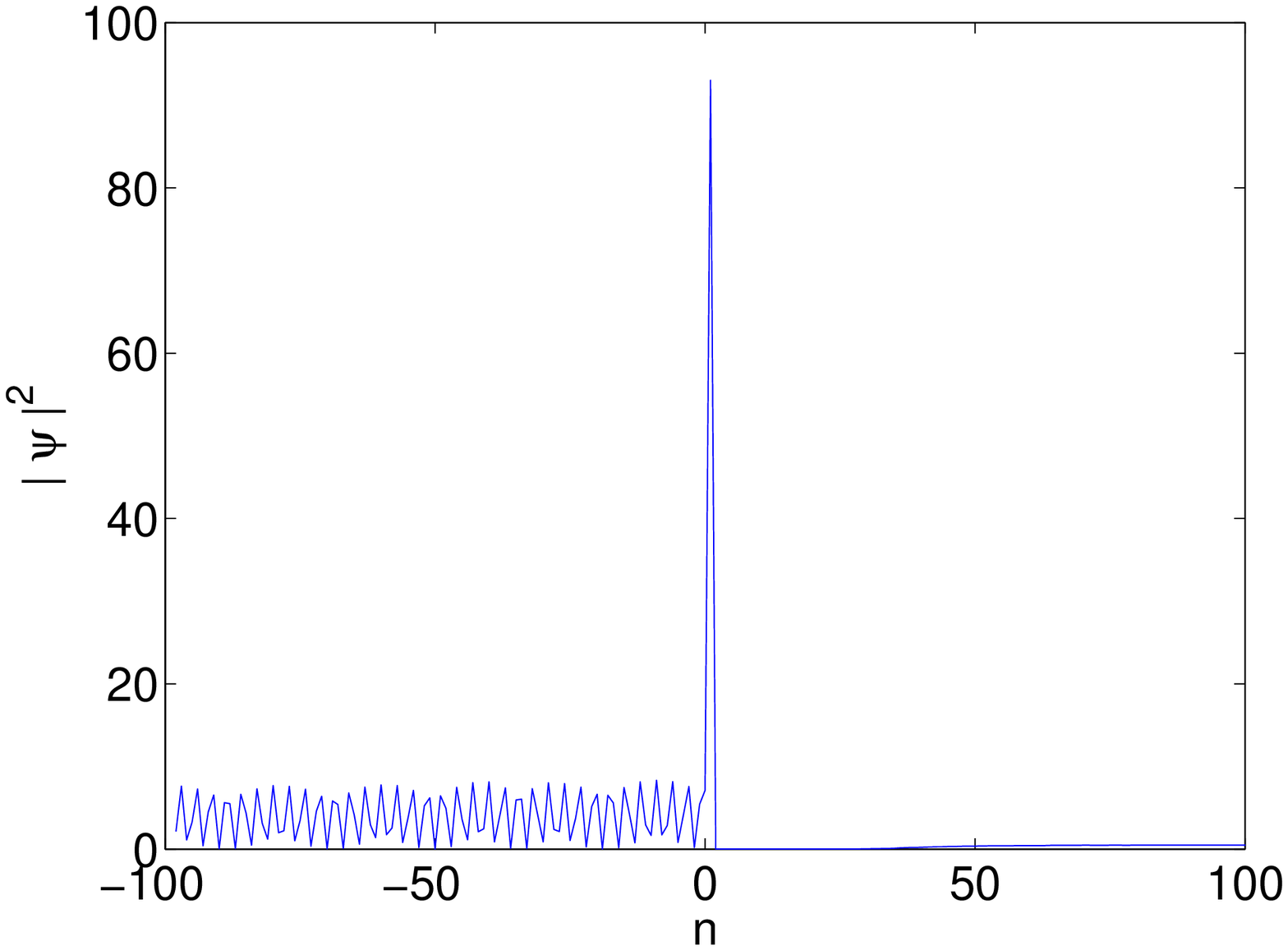}\\
\caption{ {This figure is a non-PT-symmetric analogue of Fig.~\ref{figProfiles}.  The top panel shows the contour plot of the space $n$ - time $t$ evolution of $|\phi_n|$ for the dimer (left) with $V_1=0.1i$, $V_2=-0.09i$ and $k_0=2.5$, $T=0.8$.  The trimer (right) panels have parameters $k_0=1.1, T=0.7$ with $V_1=0.1i$, $V_2=0$, $V_3=-0.09i$.  The bottom panels show individual snapshots of the solution at $t=50$ for the dimer and at $t=100$ for the trimer.} }
\label{figdeltagamma}
\end{center}
\end{figure}

\begin{figure}[tbp]
\begin{center}
\includegraphics[width=8cm,height=5.75cm]{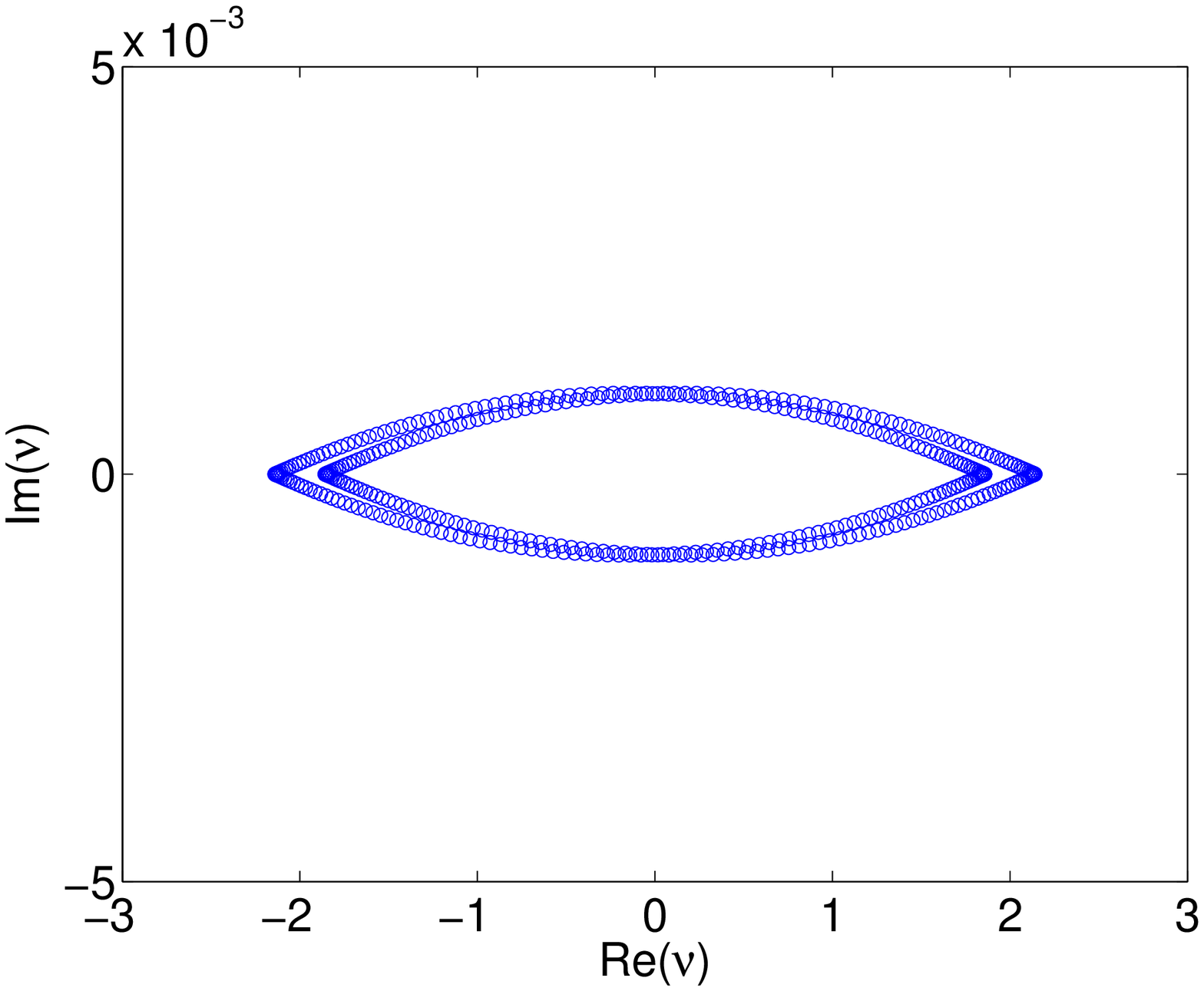}
\includegraphics[width=8cm,height=5.75cm]{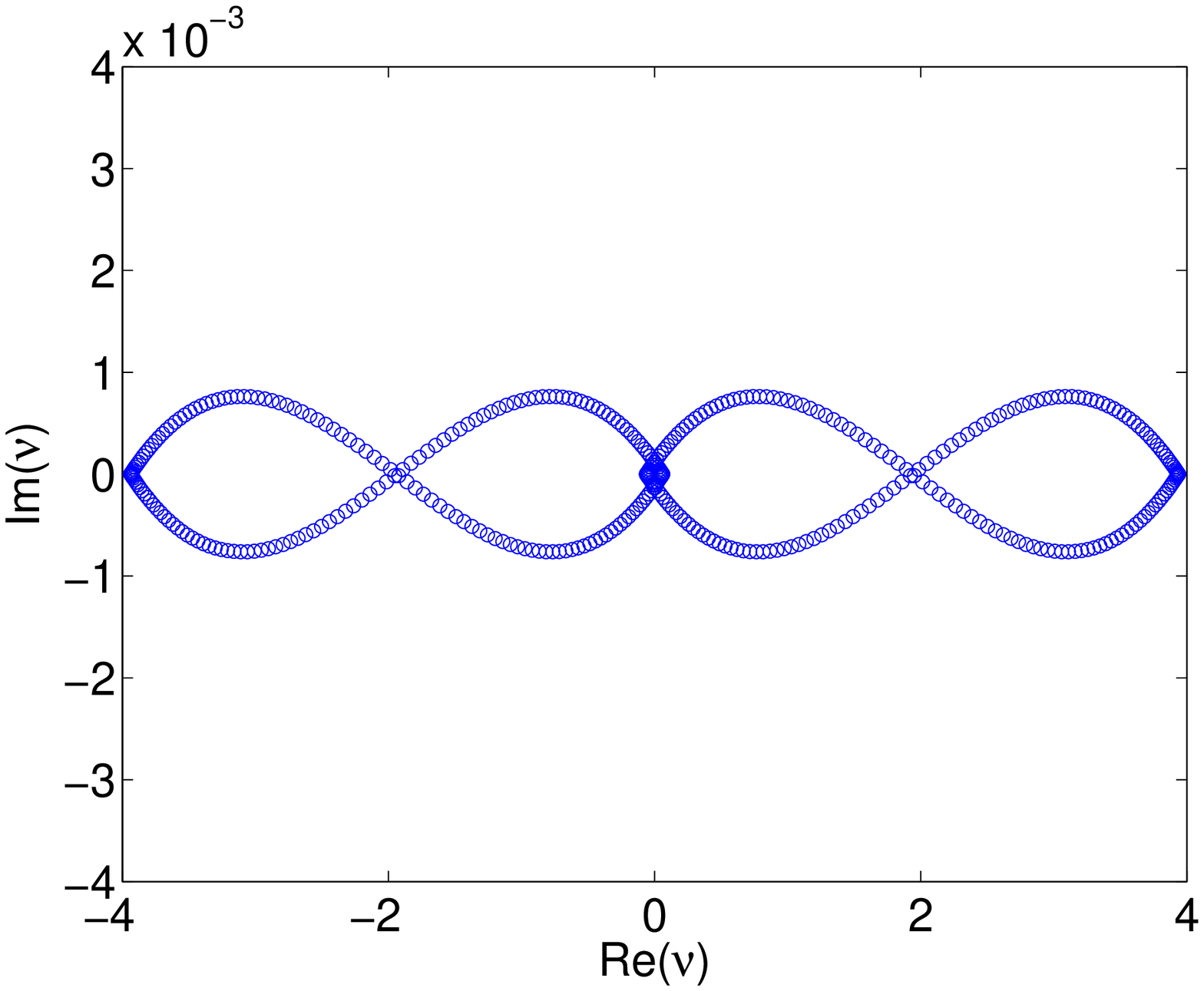}
\caption{
{
This figure is the an analogue of Fig.~\ref{figEigSpctm_add} but with an asymmetric lattice adding one extra linear node to the right-hand-side of the chain.  The panels show the eigenfrequencies for this now unstable case with the same small value of $T=0.01$ (again $k_0=1.5, \gamma=0.1$ for the dimer on the left and $k_0=0.25, \gamma=0.1$ for the trimer on the right).
}
}
\label{figxnode}
\end{center}
\end{figure}

We now turn to computations in order to quantify the above theoretical
results (as well as to extend beyond the range of what is analytically
tractable).

We start with the consideration of the transmittivity $t(k_0,T)$ and
of the rectification factor $f(k_0,T)$ which are given for the case
of the PT-symmetric 
dimer in Fig.~\ref{fig0} and for the PT-symmetric 
trimer in Fig.~\ref{fig1}. The nonlinearity is uniform in both
cases, but the linear potential is $V_1=-V_2=i \gamma$ in the former,
while it is $V_1=-V_3=i \gamma$ and $V_2=0$ in the latter.
In both cases, a typical example of the transmittivity dependence
for $k_0=2$ and $k_0=-2$ is shown in the bottom panels of the figures.
The asymmetry which is present in the top left panels between
positive and negative values of $k_0$ and which is further quantified
in the rectification factor of the top right panels clearly makes the
case for the asymmetric wave propagation in these PT-symmetric
oligomers. Although the values used here are below the PT
transition of the underlying linear oligomer system, we have ensured that
the relevant characteristic
behavior (and presented asymmetries) exist both below and above
that transition. However, as can also be inferred from the
figures, the higher the value of PT-symmetric parameter
$\gamma$, the stronger the manifestations of the asymmetric
propagation of the waves. Another relevant observation to make here
is that although the rectification factor is by construction bounded
within $[-1,1]$, the transmittivity is not bounded by unity (contrary to
what is the case in the Hamiltonian example of~\cite{lepri}).
Hence, we can observe that in all the examples of $t$ shown the
relevant factor may exceed unity thus illustrating the existence of
gain in the system. { It should be also noted that the 
rectification mechanism considered in 
Ref.~\cite{lepri} relies on multistability and resonance shifts. 
A remarkable difference here is that in the instances shown in Fig.~\ref{fig0}
there is no multistability, i.e., the output is a single-valued 
function of the input. Despite this, the rectifying effect remains
sizeable as $|f|$ can be still pretty large. This can be attributed
to the presence of gain in the systems that appears to enhance 
the transmission with respect to the Hamiltonian case. It should be noted
that similar observations can be made for the trimer in the
case of small $\gamma$ (cf. left panel of Fig.~\ref{fig1}). Yet for
sufficiently large values of $\gamma$ (cf. the right panel of 
Fig.~\ref{fig1}) multistability arises and, in turn, contributes
to the observed rectification.}

%To propagate the extended solutions $\psi_n$ found in Section \ref{secstatDNLS} we apply the perturbation theory of Section \ref{secTDDNLS} with $\Lambda=-\omega=\cos(k)$.  Again the solutions are propagated with a fourth order Runge-Kutta algorithm according to (\ref{eq: TDDNLS}).  We find that the extended solutions are unstable, see Figures \ref{figEigSpctm} and \ref{figProfiles}.

{ We now wish to touch upon the problem of the stability of the 
extended solutions
$\psi_n$, identified in the form of Eq.~(\ref{eq: FormOfPsi_n}). 
%in Section \ref{secstatDNLS}.
A complete stability analysis would 
require to take into account the fact that the lattice is infinite and that
the eigemodes associated with the unstable eigenvalues (when they exist) 
are exponentially localized around the oligomers. To our knowledge an
exhaustive study of this problem has not been reported in the literature
(see e.g. Ref.~\cite{malaz} for a related study in the continuum case). 
%Of course, for small nonlinearities/amplitudes the solutions should be
%stable but is of little relevance for the type 
%of nonreciprocal solutions we are interested in, which are strongly nonlinear.
A more complete analysis will be reported elsewhere.
In the present work we limit ourselves to the illustration of
a few representative cases.
Typically, the relevant solutions 
are found to be unstable. An example of the corresponding spectral
plane (Re$(\nu)$,Im$(\nu)$) of eigenfrequencies $\nu=$Re$(\nu)$ +i Im$(\nu)$
is shown for the case of the dimer and a corresponding
one of the trimer in Fig.~\ref{figEigSpctm}. }
%The right panels of the figures
%illustrate the dynamical evolution of the system initiated by means
%of an unstable
In fact, our computations indicate that this feature persists in the
Hamiltonian variant of the model of Ref.~\cite{lepri} 
(although a more detailed examination
of the latter is of interest in its own right). The dynamical instability
observed via the presence of imaginary eigenfrequencies in the spectral
plane of the PT-symmetric oligomers is corroborated by direct numerical
simulations in Fig.~\ref{figProfiles}. Here, we initialize the lattice
with the solutions that were shown to be unstable via the spectral
analysis of Fig.~\ref{figEigSpctm}, to which a normally distributed small
amplitude random perturbation has been superposed. The manifestation
of the instability once again clearly underscores the role of the 
gain in the system. In particular, the dynamics reveals the tendency
of the site which has gain to acquire a super-critical ``mass'' (or
density), which subsequently grows exponentially well beyond the
density of the linear background (or of the sites with loss).

{ A natural question about these extended states concerns the
dependence of the instability on the strength of the nonlinearity.
We explore this issue now by means of the results shown in 
Fig.~\ref{figEigSpctm_add}. In particular, we vary the value of
the transmitted amplitude $T$ in order to systematically
approach the linear limit which corresponds to $T \rightarrow 0$.
 For small $T$, the overall amplitude of the
solution is small and the nonlinearity within the problem becomes negligible,
giving essentially rise to linear states. For such sufficiently
small values of $T$, the growth rate of the relevant instability is found
to be negligible.
We have given an example in the figure of such a small T, 
stable waveform
and its linear stability analysis, and also show the dependence of the growth
rate of the instability through 
the largest negative imaginary eigenfrequency as a function of T both
for the dimer and the trimer. The latter clearly illustrates how the
instability strength grows as a function of the nonlinearity.  
For both the dimer and the trimer solutions instability arises
roughly at  $T\approx 0.25$, as we see in 
the bottom panel of Fig.~\ref{figEigSpctm_add}.
}

Given the above considerations and also the lesser physical relevance
of exciting an initial condition of the form of 
Eq.~(\ref{eq: FormOfPsi_n}) in an optical setting (which would be
the prototypical potential realization of the considerations presented
herein), we now turn to the examination of the dynamics of a 
more localized Gaussian
wavepacket. The latter is assumed to be of the form: 
\begin{equation}
\psi_n(0)=I e^{ik_0 n- {(n-n_0)^2}/{s^2}},
\end{equation}
i.e., of amplitude $I$, 
centered at $n_0\in\mathds{Z}$, while its speed is controlled
by the parameter $k_0$ { 
($k_0>0, n_0<0$ for left-incoming packets and
$k_0<0, n_0>N$ for right-incoming ones)}. The value of $s$ 
(chosen to assume the value $10$ in what follows) determines the wavepacket
width. { To minimize the dispersive effects and thus render
negligible the dependence 
of the scattering process on the initial position $n_0$, we limited ourselves to
the case in which $|k_0|=\pi/2$. It should, however, be noted that
for different values of $k_0$ (and especially for ones significantly
deviating from the above band center), the dispersive nature of the discrete
medium would make the results dependent not only on $k_0$ but also $n_0$.} 
It should be noted here that this aspect of our consideration
is purely numerical (as it is not straightforward to obtain explicit 
analytical expressions for the transmission and reflection coefficients
in this case). { For a nonlinear system, there is no straightforward 
correspondence between the transmission coefficients of plane waves 
and the one of a wavepacket (in the language of Ref.~\cite{knapp} the latter it is 
a fixed input problem).
Nevertheless, arguably, both problems are
of interest. The fixed output problem  examined above
can be analytically studied
and features such as bistability or asymmetric transmittivity can be quantified.
Yet in the fixed input problem considered
below some of these features (such as the asymmetric 
transmittivity) can be measured in ways that are conducive towards 
experimental realizations.
} We thus evolve an incident wavepacket from the left or
from the right and observe its propagation 
as illustrated in the space-time contour plots of Fig.~\ref{figGauss}.
In what is shown below, we have performed the numerical computations
for $I=1$, but we have confirmed that the relevant phenomenology
persists for a wide range of non-vanishing values of $I$ (approaching
the linear limit of $I \rightarrow 0$ leads to vanishing transmittivity
differences, once again due to the principle of reciprocity).
%was initiated on the lattice and propagated with a fourth order Runge-Kutta to obtain $\phi_n(t)$ satisfying (\ref{eq: TDDNLS}).  One sees the effect of the nonlinear nodes in Figure \ref{figGauss} as the wave partially transmits and reflects.

The transmission coefficient in this case is defined as the fraction of the 
total sum of the squared modulus of the field across nodes
that is transmitted to the right of the PT-symmetric
oligomer after the Gaussian wavepacket passes over the relevant nodes.  
That is, we define $t_{k_0}=\sum_{n<1}|\phi_n|^2/\sum_n|\phi_n|^2$ for 
$k_0<0$ and $t_{k_0}=\sum_{n>N}|\phi_n|^2/\sum_n|\phi_n|^2$ for $k_0>0$.  
We find that the transmittivity is greater for Gaussians approaching from the 
right and that the  transmittivity difference is amplified as $\gamma$ 
increases, see Fig. \ref{figtdiff} which presents the relevant 
rectifying factor. This can be qualitatively understood once again on the
basis of the structure of the PT oligomer. The node that has gain (which
is on the left) favors reflection for a wavepacket from the left, while
it favors transmission for a wavepacket impinging from the right, hence
ensuring that $t_{-k_0}>t_{k_0}$ in our setup.  
Interestingly, in this case of the Gaussian wavepacket as well, 
for a high enough value of $\gamma$ the density  accumulates and
grows indefinitely at 
the nonlinear node that bears the gain.   On the dimer { with $\alpha_1=\alpha_2=1$} we find that for $k_0=\pi/2$  this critical value is $\gamma\simeq 0.6531$, and for $k_0=-\pi/2$ it is 
$\gamma\simeq 1.0890$.  On the trimer with { with $\alpha_1=\alpha_2=\alpha_3=1$} $k_0=\pi/2$ the blowup occurs for $\gamma\simeq 0.6737$ and with $k_0=-\pi/2$ the value is $\gamma\simeq 0.8043$.  We note once again that these features are unique to the case of PT-symmetric
oligomers through their gain-loss pattern and would be entirely
absent in the earlier Hamiltonian installment of such chains in~\cite{lepri}.
{ This phenomenon can be seen as an illustration for the present
system (and considered input wavepackets) 
of the phenomenology associated with the
PT phase transition and its nonlinear analogs considered e.g. in works
such as Refs.~\cite{kot1,pgk} among others. What was found in the above
systems (and is manifest here as well) is that past a critical value
of gain-loss, the oligomer settings (and their lattice embedding generalization)
are unable to balance the intended (through the gain-loss) growth and
decay features through a gradient of the phase of the configuration.
Instead, they become subject to indefinite growth (and corresponding decay
in the lossy site). It should also be added here that Fig.~\ref{figtdiff}
additionally contains the possibility of asymmetric nonlinearities
(in the spirit of~\cite{lepri}), with the two sites bearing
$\alpha_1=1-\Delta\alpha/2, \alpha_2=1+\Delta\alpha/2$ and the three sites
bearing 
$\alpha_1=1-\Delta\alpha/2, \alpha_2=1, \alpha_3=1+\Delta\alpha/2$ 
for the dimer and trimer, respectively. It can be seen that the latter
asymmetry provides a smooth change of the rectification 
factor with respect to the case $\Delta \alpha=0$, which is more pronounced
the higher the value of $\Delta \alpha$.}

{ As a final comment, we revisit the comparison with the results
of Ref.~\cite{lepri} and touch upon the specific role of the gain-loss
balance within the PT-symmetric settings considered herein.
As regards the former, and as alluded to previously, 
the broken parity symmetry is essential to the 
phenomenology reported herein and even to that of Ref.~\cite{lepri}. 
However, although
in the latter the phenomenology was chiefly due to resonance shifts and 
creation
of multi-stability, here it is far more so due to the existence of 
loss and gain. As we also illustrated in Fig.~\ref{figtdiff},
by breaking the parity
of the system within its nonlinear term, we may weakly affect the
relevant phenomenology (or more strongly depending on the size
of the parity breaking), yet the fundamental effects are still present
and dominated by the loss-gain pattern.

Another deep question concerns the specific role
of the gain-loss balance within our PT-symmetric system.
In a sense, the exact pattern of gain-loss and its perfect balance
is not that important for some aspects of the observed phenomenology. 
What is more important is that there be a
parity breaking through the gain-loss pattern which will, in turn,
induce the relevant transmission asymmetries and also features particular
to our system (and not to the Hamiltonian variant of Ref.~\cite{lepri})
such as the indefinite growth and the transition to it beyond a certain
gain strength. This is evident in Fig.~\ref{figdeltagamma} which showcases
the instability induced growth for a case where the gain and loss 
coefficients are not perfectly balanced.
To some extent, it should be added that something of this sort may be
expected from past experience with variants of PT systems and their 
features, including e.g. the  observation of the passive-PT phase 
transition~\cite{salamo},
 where the system only had two sites one without gain-loss 
and the other only with loss.
Nevertheless, it should be added that there {\it are} features of the
system that critically (and quite subtly) may depend on whether the gain-loss
pattern is perfectly balanced or not. We have accidentally bumped into
a very simple and instructive example of this kind, when considering
the effect of the lattice size. Consider a dimer embedded within the
sites 100 and 101 of a lattice of size $M=200$. Now, consider the same
dimer within a lattice of size $M=201$. It is straightforward to see
that the former is perfectly PT-symmetric, while the latter is not
(by what appears to be a negligible --far away-- boundary contribution).
Yet, if one considers the spectrum of the linear problem (and by
extension the linearization spectrum of nonlinear states) within the
two cases, one will {\it immediately} observe that the former system
will be stable up to a finite critical point $\gamma_{cr}$, while the
latter will become immediately (and nontrivially) unstable due to its
violation of the exact PT symmetry -- compare the spectra in 
Figs.~\ref{figEigSpctm_add} and \ref{figxnode}.   
Moreover, the stronger and immediate instability implied by 
Fig.~\ref{figxnode} has a bearing on the dynamics of the kind
shown in Fig.~\ref{figdeltagamma} (cf. with the case of
Fig.~\ref{figProfiles}) given that the ensuing growth is manifested
for earlier times in the case of asymmetric gain-loss.
Hence, while certain 
aspects of the phenomenology (like the transmission asymmetry in the dynamics)
may not be critically hinging on the perfect balance of gain and
loss, others such as the stability of the nonlinear states (or the
spectrum of the linear ones) upon introduction of gain and loss
clearly do.}

\section{Conclusions}

In the present work, we considered a lattice setting where embedded
in a linear Schr{\"o}dinger chain was a nonlinear PT-symmetric
oligomer, typically a dimer or a trimer. Our analytical considerations
were focused around plane waves enabling us to analytically compute
both the transmittivity and the rectification factor between left-
and right-propagating such waves. These features amply evidenced
the asymmetric nature of the propagation and even illustrated features
particular to the gain-loss systems, such as the existence of
over-unity transmittivities. On the other hand, we also considered
the spectral stability of such states revealing { their typical
instability (except for the near-linear case) }
that was monitored and shown dynamically to lead to
mass focusing on a single (gain) node of the lattice. This, in turn,
led us to the consideration of the asymmetry of propagation
of a Gaussian wavepacket which we numerically quantified. Here, too,
however the interesting phenomenon of the potential trapping of mass
in a particular site with gain and the subsequent indefinite growth
thereof were observed and quantified. { 
The effect of asymmetries in the nonlinearity and in the gain/loss
profile were also examined.}

These results suggest numerous interesting investigations for future
work. It would be relevant to attempt to theoretically
quantify transmittivities
and reflectivities of a Gaussian wavepacket perhaps through a 
judicious variational ansatz or some similar method that 
appropriately reduces the degrees of freedom, while taking into consideration
both the complex PT-oligomer dynamics and the generic
existence of a reflecting and a transmitting wavepacket.
On the other hand, it would be particularly interesting to
generalize the relevant considerations to higher dimensional
settings and examine how incident waves from different directions
may affect the transmittivity of different forms of two-dimensional
PT-symmetric oligomers (in the simplest genuinely two dimensional
case, PT-symmetric squares~\cite{guenther}). These themes will be considered
in future studies.

%\begin{acknowledgments}

\acknowledgments 
PGK gratefully acknowledges support from the US-NSF through grants 
DMS-0806762 and CMMI-1000337 and
from the Alexander von Humboldt Foundation, the Alexander
S. Onassis Public Benefit Foundation (through grant RZG 003/2010-2011)
and from the Binational Science Foundation (through grant 2010239).
SL acknowledges support from the Miur PRIN 2008
project \textit{Efficienza delle macchine termoelettriche:
un approccio microscopico}.

%\end{acknowledgments}

\appendix
\section{Appendix}

For the record we show four iterations of (\ref{eq: alg}) starting from $\Psi_{N+1} = e^{ik}$ and $\Psi_N = 1$:
\begin{eqnarray}
\label{eq: algN}
\delta_{N} &=& V_{N}-\omega+\alpha_{N}|T|^2  \nonumber\\
\Psi_{N-1} &=& - \Psi_{N+1}+\delta_N\Psi_N         \qquad (l=1)\nonumber\\
&=& - e^{ik}+\delta_N\nonumber\\
\delta_{N-1} &=& V_{N-1}-\omega+\alpha_{N-1}|T|^2|\delta_N-e^{ik}|^2\nonumber\\
\Psi_{N-2} &=& - \Psi_{N}+\delta_{N-1}\Psi_{N-1}         \qquad (l=2)\nonumber\\
&=& - 1 + \delta_{N-1}(\delta_N-e^{ik})\nonumber\\
\delta_{N-2} &=& V_{N-2}-\omega+\alpha_{N-2}|T|^2| 1 + \delta_{N-1}(e^{ik}-\delta_N)|^2\nonumber\\
\Psi_{N-3} &=& - \Psi_{N-1}+\delta_{N-2}\Psi_{N-2}         \qquad (l=3)\nonumber\\
&=& e^{ik} - \delta_N+\delta_{N-2} \left(   - 1 + \delta_{N-1}(\delta_N-e^{ik})  \right)\nonumber\\
&=& -\delta_{N-2}+(e^{ik}-\delta_N)(1-\delta_{N-2}\delta_{N-1})\nonumber\\
\delta_{N-3} &=& V_{N-3}-\omega+\alpha_{N-3}|T|^2|\delta_{N-2}+(\delta_N-e^{ik})(1-\delta_{N-2}\delta_{N-1})|^2 \nonumber\\
\Psi_{N-4} &=& - \Psi_{N-2}+\delta_{N-3}\Psi_{N-3}         \qquad (l=4)\nonumber\\
&=&  1 + \delta_{N-1}(e^{ik}-\delta_N)+\delta_{N-3}\left(   -\delta_{N-2}+(e^{ik}-\delta_N)(1-\delta_{N-2}\delta_{N-1})      \right)  \nonumber\\
&=& 1 -\delta_{N-3}\delta_{N-2}  + (e^{ik}-\delta_N)\left(   \delta_{N-1}+\delta_{N-3} (1-\delta_{N-2}\delta_{N-1})      \right)\nonumber\\
\delta_{N-4} &=& V_{N-4}-\omega+\alpha_{N-4}|T|^2 |  1 -\delta_{N-3}\delta_{N-2}  + (e^{ik}-\delta_N)\left(   \delta_{N-1}+\delta_{N-3} (1-\delta_{N-2}\delta_{N-1})      \right)|^2\nonumber
%\\
%\Psi_{N-5} &=& - \Psi_{N-3}+\delta_{N-4}\Psi_{N-4}         \qquad (l=5)\nonumbe%r\\
%&=& \delta_{N-2} - (e^{ik}-\delta_N)(1-\delta_{N-2}\delta_{N-1})  \nonumber\\
%&& \qquad\qquad + \delta_{N-4}\left(  1 -\delta_{N-3}\delta_{N-2}  + (e^{ik}-\d%elta_N)\left(   \delta_{N-1}+\delta_{N-3} (1-\delta_{N-2}\delta_{N-1})      \ri%ght)   \right)\nonumber\\
%&=& \delta_{N-2}+\delta_{N-4}(1-\delta_{N-3}\delta_{N-2})\nonumber\\
%&&\qquad\qquad + (e^{ik}-\delta_N)\left(  \delta_{N-4}\delta_{N-1} + (\delta_{N%-4}\delta_{N-3}-1)(1-\delta_{N-2}\delta_{N-1}) \right).\nonumber
\end{eqnarray}
Notice that if we iterate $N$ times we get $\delta$'s equal to $- \omega$ since $V=\alpha=0$ for these nodes.  Then $R_0$ can be calculated from (\ref{eq: R0RwrtPsi}) in terms of appropriate $\delta$'s.

For $N=3$, the algorithm gives a transmission coefficient
\begin{equation}
t=\left|  \frac{e^{ik}-e^{-ik}}{      e^{ik}-\delta_1 + (e^{ik}-\delta_3)(1-\delta_2(\delta_1-e^{ik}) )                }  \right|^2
\end{equation}
for
\begin{eqnarray}
\delta_3 &=& V_3-\omega+\alpha_3|T|^2\nonumber\\
\delta_2 &=& V_2-\omega+\alpha_2 |T|^2 |\delta_3-e^{ik}|^2 \nonumber\\
\delta_1 &=& V_1-\omega+\alpha_1 |T|^2|1+\delta_2(e^{ik}-\delta_3)|^2.
\end{eqnarray}
Similar results can be obtained for $N=4, 5, \dots$.


\begin{thebibliography}{99}



\bibitem{bend} C.M. Bender and S. Boettcher,
Phys. Rev. Lett. {\bf 80}, 5243 (1998); C.M. Bender, S. Boettcher
and P.N. Meisinger, J. Math. Phys. {\bf 40}, 2201 (1999)
%math-ph/9712001
%quant-ph/9809072

%;C.M. Bender, Rep. Prog. Phys. {\bf 70}, 947 (2007).
%astro-ph/0704.2291   (?)
\bibitem{muga} A. Ruschhaupt, F. Delgado, F. and J.G. Muga,
J. Phys. A: Math. Gen., {\bf 38}, L171 (2005); 
A. Mostafazadeh and F. Loran, EPL, {\bf 81} 10007 (2008). 

\bibitem{christo1} Z.H. Musslimani, K.G. Makris, R. El-Ganainy
and D.N. Christodoulides, Phys. Rev. Lett. {\bf 100}, 030402 (2008);
K.G. Makris, R. El-Ganainy, D.N. Christodoulides and Z.H. Musslimani,
Phys. Rev. A {\bf 81}, 063807 (2010).
%http://prl.aps.org/abstract/PRL/v100/i3/e030402
%http://pra.aps.org/abstract/PRA/v81/i6/e063807

\bibitem{salamo} A. Guo, G. J. Salamo, D. Duchesne, R. Morandotti,
  M. Volatier-Ravat, V. Aimez,
G. A. Siviloglou and D. N. Christodoulides,
  Phys. Rev. Lett. {\bf 103}, 093902 (2009).
  %http://prl.aps.org/abstract/PRL/v103/i9/e093902


\bibitem{kip} C.E. R{\"u}ter, K.G. Makris, R. El-Ganainy,
D.N. Christodoulides, M. Segev, D. Kip,
Nature Phys. {\bf 6}, 192 (2010).
%http://www.nature.com/nphys/journal/v6/n3/pdf/nphys1515.pdf

\bibitem{tsampikos_recent} J. Schindler, 
A. Li, M.C. Zheng, F.M. Ellis and T. Kottos,
Phys. Rev. A {\bf 84}, 040101 (2011).
%http://pra.aps.org/abstract/PRA/v84/i4/e040101

\bibitem{kot1} H. Ramezani, T. Kottos, R. El-Ganainy and D.N.
Christodoulides, Phys. Rev. A {\bf 82}, 043803 (2010).
%optics/1005.5189

\bibitem{sukh1} A.A. Sukhorukov, Z. Xu and Yu.S. Kivshar,
Phys. Rev. A {\bf 82}, 043818 (2010).
%optics/1009.5428

\bibitem{kot2} M.C. Zheng, D.N. Christodoulides, R. Fleischmann
and T. Kottos, Phys. Rev. A {\bf 82}, 010103(R) (2010).
%http://pra.aps.org/abstract/PRA/v82/i1/e010103

\bibitem{grae1} E.M. Graefe, H.J. Korsch and A.E. Niederle,
Phys. Rev. Lett. {\bf 101}, 150408 (2008).
%quant-ph/0807.1777

\bibitem{grae2} E.M. Graefe, H.J. Korsch and A.E. Niederle,
Phys. Rev. A {\bf 82}, 013629 (2010).
%quant-ph/1003.3355

\bibitem{kot3} Z. Lin, H. Ramezani, T. Eichelkraut, T. Kottos,
H. Cao and D.N. Christodoulides, Phys. Rev. Lett. {\bf 106}, 213901 (2011).
%optics/1108.2493

\bibitem{pgk} K. Li and P. G. Kevrekidis
Phys. Rev. E {\bf 83}, 066608 (2011)
%nlin/1102.0809

\bibitem{dmitriev1} S.V. Dmitriev, S.V. Suchkov, A.A. Sukhorukov, 
and Yu.S. Kivshar,
Phys. Rev. A {\bf 84}, 013833 (2011) 
%http://pra.aps.org/abstract/PRA/v84/i1/e013833

\bibitem{dmitriev2} S.V. Suchkov,  B.A. Malomed, S.V. Dmitriev and 
Yu.S. Kivshar, Phys. Rev. E 84, 046609 (2011). 
%http://pre.aps.org/abstract/PRE/v84/i4/e046609

\bibitem{miron} A.E. Miroshnichenko, B.A. Malomed, and Yu.S. Kivshar
Phys. Rev. A {\bf 84}, 012123 (2011).
%math-ph/1104.0849

\bibitem{konorecent} F.Kh. Abdullaev, Y.V. Kartashov, V.V. Konotop
and D.A. Zezyulin, Phys. Rev. A {\bf 83}, 041805 (2011)
%nlin/1104.0276

\bibitem{konorecent2} D. A. Zezyulin, Y. V. Kartashov, V. V. Konotop,
arXiv:1111.0898. %nlin

\bibitem{kosevich} Yu. A. Kosevich, Phys. Rev. B, 
{\bf 52}, 1017 (1995).

\bibitem{l1}  M. Terraneo, M. Peyrard, and G. Casati, Phys. Rev. Lett.
{\bf 88}, 094302 (2002).
%cond-mat/0201125

\bibitem{l3} C.W. Chang, D. Okawa, A. Majumdar, and A. Zettl,
Science {\bf 314}, 1121 (2006); W. Kobayashi, 
Y. Teraoka, and I. Terasaki, Appl. Phys.
Lett. {\bf 95}, 171905 (2009).
%http://www.sciencemag.org/content/314/5802/1121.abstract
%cond-mat/0910.1153

\bibitem{l5} M. Scalora, J. P. Dowling, C. M. Bowden, and M. J.
Bloemer, J. Appl. Phys. {\bf 76}, 2023 (1994);
M. D. Tocci, M. J. Bloemer, M. Scalora, J. P. Dowling, and
C. M. Bowden, Appl. Phys. Lett. {\bf 66}, 2324 (1995).
%http://jap.aip.org/resource/1/japiau/v76/i4/p2023_s1?isAuthorized=no
%http://apl.aip.org/resource/1/applab/v66/i18/p2324_s1?isAuthorized=no


\bibitem{l7} V.V. Konotop and V. Kuzmiak, Phys. Rev. B {\bf 66}, 235208
(2002). B. Liang, B. Yuan and J.C. Cheng, Phys. Rev. Lett. {\bf 103},
104301 (2009).
%http://prb.aps.org/abstract/PRB/v66/i23/e235208
%http://prl.aps.org/abstract/PRL/v103/i10/e104301

\bibitem{l9} K. Gallo, G. Assanto, K. Parameswaran and M. Fejer,
Appl. Phys. Lett. {\bf 79}, 314 (2001).
%http://apl.aip.org/resource/1/applab/v79/i3/p314_s1?isAuthorized=no

\bibitem{l10} M.W. Feise, I.V. Shadrivov and Y.S. Kivshar,
Phys. Rev. E {\bf 71}, 037602 (2005).
%optics/0412009

\bibitem{chiara1} N. Boechler, G. Theocharis and C. Daraio,
Nature Materials {\bf 10}, 665 (2011).
%http://www.nature.com/nmat/journal/v10/n9/abs/nmat3072.html

\bibitem{others} B. Liang, B. Yuan and J.C. Cheng, 
%Acoustic diode: Rectification of acoustic
%energy flux in one-dimensional systems. 
Phys. Rev. Lett. 103, 104301 (2009);
B. Liang, X.S. Guo, J. Tu, D. Zhang and J.C. Cheng, 
%An acoustic rectifier.
Nature Mater. {\bf 9}, 989992 (2010).
%http://prl.aps.org/abstract/PRL/v103/i10/e104301
%http://www.nature.com/nmat/journal/v9/n12/full/nmat2881.html


\bibitem{longhi} S. Longhi,
J. Phys. A. {\bf 44}, 485302 (2011).
%quant-ph/1111.3448

\bibitem{lepri} S. Lepri and G. Casati, Phys. Rev. Lett.
{\bf 106}, 164101 (2011).
%nlin/1103.4802

\bibitem{vvk} D. A. Zezyulin, V. V. Konotop,
arXiv:1202.3652.

\bibitem{tsir1} G.P. Tsironis, M.I. Molina and D. Hennig,
Phys. Rev. E {\bf 50}, 2365 (1994).

\bibitem{moli} M.I. Molina, H. Bahlouli
Phys. Lett. A {\bf 284}, 87 (2002). 

\bibitem{moli2} M. I. Molina
Phys. Rev. B {\bf 60}, 2276 (1999). 

\bibitem{efrem} K. Hizanidis, Y. Kominis and N.K. Efremidis,
Opt. Express {\bf 16}, 18296 (2008).


\bibitem{knapp} R. Knapp, G. Papanicolaou and B. White,
J. Stat. Phys. 63, 567, (1991).

\bibitem{malaz} B.~A. Malomed and M.~Ya. Azbel.
Phys. Rev. B, {\bf 47}, 10402--10406 (1993).


\bibitem{lindquist} B.Lindquist, Phys. Rev. E
\textbf{63}, 56605 (2001).

\bibitem{guenther} K. Li, P.G. Kevrekidis, B.A. Malomed,
U. Guenther, arXiv:1204.5530.

\end{thebibliography}
\end{document}